\documentclass[useAMS,usenatbib,usegraphicx,onecolumn]{mn2e} 

\title[statistics of NIRS0S]{Statistics of the structure components in S0s: implications for bar induced secular evolution}

\author[E-mail:eija.laurikainen@oulu.fi]{E. Laurikainen$^{1,2}$\thanks{E-mail:eija.laurikainen@oulu.fi}, H. Salo$^{1}$, E. Athanassoula$^{4}$, A. Bosma$^{4}$, R. Buta$^{3}$, J. Janz$^{1,5}$ \\
$^{1}$Dept. of Physics/Astronomy Division, University of Oulu, FI-90014 Finland\\
$^{2}$Finnish Centre of Astronomy with ESO (FINCA), University of Turku, V\"ais\"al\"antie 20, FI-21500, Piikki\"o, Finland\\ 
$^{3}$Department of Physics and Astronomy, University of Alabama, Box 870324, Tuscaloosa, AL 35487\\
$^{4}$Aix Marseille Universit\'e, CNRS, LAM (Laboratoire d'Astrophysique de Marseille) UMR 7326, 13388, Marseille, France\\
$^{5}$Astronomisches Rechen-Institut, Zentrum f\"ur Astronomie der Universit\"at Heidelberg, M\"onchhofstrase 12-14. D-69120 Heidelberg}

\onecolumn

\begin{document}

\date{Accepted: 24.1.2013}


\maketitle

\label{firstpage}

\begin{abstract}

The fractions and dimension of bars, rings and lenses are studied in
the Near-IR S0 galaxy Survey (NIRS0S), which is a sample of $\sim$200
early-type disk galaxies, mainly S0s.  We find evidence that multiple
lenses in some barred S0s are related to bar resonances in a similar
manner as the inner and outer rings, for which the outer/inner length
ratio $\sim$ 2. Inner lenses in the non-barred galaxies normalized to
galaxy diameter are clearly smaller than those in the barred
systems. Interestingly, these small lenses in the non-barred galaxies
have similar sizes as barlenses (lens-like structures embedded in a
bar), and therefore might actually be barlenses in former barred
galaxies, in which the outer, more elongated bar component, has been
destroyed. We also find that fully developed inner lenses are on
average a factor 1.3 larger than bars, whereas inner rings have
similar sizes as bars. The fraction of inner lenses is
found to be constant in all family classes (A, AB, B).  

Nuclear bars appear most frequently among the weakly barred (AB)
galaxies, which is consistent with the theoretical models by
Maciejewski $\&$ Athanassoula (2008). Similar sized bars as the
nuclear bars were detected in seven 'non-barred' S0s.  Galaxy
luminosity does not uniquely define the sizes of bars or bar-related
structures, neither is there any upper limit in galaxy luminosity for
bar formation. Although all the family classes cover the same range of
galaxy luminosity, the non-barred (A) galaxies are on average 0.6 mag
brighter than the strongly barred (B) systems.

Overall, our results are consistent with the idea that bars play an
important role in the formation of the structure components of
galaxies. The fact that multiple lenses are common in S0s, and that at
least the inner lenses can have very old stellar populations, implies
that the last destructive merger, or major gas accretion event, must
have taken place at a fairly high redshift.

\end{abstract}


\section{Introduction}

Galaxy morphology is a subject of significant recent turmoil, both
theoretically and observationally. The early-type disk galaxies have
an important role in this debate. In the current paradigm of galaxy
formation \citep{white1978,frenk1985}
morphology at
high redshifts is a transient property changing between spirals and
ellipticals, e.g disks are renovated after each merger event, unless
the gas in the halo becomes too hot for participating in the disk re-building 
anymore. After the merger dominated era, intrinsic secular dynamical
evolution is expected to play an increasing role.  A 
different view is taken by \citet{sales2011} 
who suggested that most stars in bulges and disks
form in situ from hot and cold gas falling in from the halo. 
In that case no correlation is expected between the stellar population
ages and the epochs of the formation of structures in galaxies.

Even within the standard picture of galaxy evolution there is an active
debate of how important secular evolution is in the formation of structures
of local galaxies, when did the secular processes start to
dominate, and what are the driving forces of that
evolution. It has been suggested \citep{hammer2009,puech2012} 
that even up to 50$\%$ of the spiral galaxy disks might have been formed via disk
rebuilding soon after the most recent merger event. On the other hand,
dynamical evolution can be largely driven by bars \citep{sellwood1993,kormendy2004,hopkins2010,atha2012}, without invoking any significant gas accretion.
One important problem in the present-day cosmological simulations is 
that disk galaxies with excessive
$B/T$ flux-ratios and too large values of the S\'ersic index are created \citep{navarro1991},
unless the
galaxies have extremely quiescent merger and gas accretion histories
\citep{martig2012,kraljic2011}.
Disk-like bulges (low S\'ersic index) are created also in cosmological simulations
by starbursts at high redshifts,
but those bulges are 
still very massive compared to typically observed bulges \citep{wang2012,okamoto2012}.
In non-cosmological simulations, where it is possible to study a controlled
range of various parameters, the problem of massive bulges can be avoided if
the galaxies evolve in very dense environments \citep{Khochfar2011},
or if two gas rich pure disk galaxies are made to merge \citep{keselman2012}. 
However, a large majority
of the S0s in the local Universe have low $B/T$ flux-ratios; moreover
their S\'ersic indexes are close to 1-2 indicating that the bulges 
are fairly exponential 
\citep{lauri2005,lauri2010}. 
In cosmological simulations disks are easily destroyed
by infalling satellites \citep{toth1992},
or by accretion
of misaligned gas \citep{scanna2009,sales2011},
which is challenging when explaining the observed large-scale
lenses in many S0 galaxies.

Lower $B/T$ flux-ratios and longer lasting disk structures are
expected if galaxy evolution after z$\sim$1 was mostly quiescent.
In that case the S0s could be largely red and dead 
(or at most partly re-rejuvenated) former spirals in which the gas does not
participate in the disk re-building anymore.  Strong support for
this comes from the recently suggested scenario, in which the S0s form
a parallel sequence with the spirals in the Hubble diagram
\citep{lauri2010,cappellari2011,kormendy2012},
renovating the old idea of \citet{vandenbergh1976}. 
In that scenario both the early and late-type spirals evolve directly
into S0s forming S0a, S0b and S0c types. 
This makes it possible to understand the observed small $B/T$ flux-ratios
in some S0s,
and also why the properties of the bulges and disks are so
similar in the S0s and spirals \citep{lauri2010}. 
Motivated by the fact that as much as 70$\%$ of the S0s live in galaxy groups 
\citep{willman2009,just2011}
\citet{eliche2012}  suggested 
that the transformation from spirals into S0s could occur
via minor mergers, leading to the observed parallel sequence, provided 
that the bulges were originally very small disk-like components.  Bars 
might also be a driving force in the transformation
process, though that has not been studied in detail yet.

Understanding the formation of bars, lenses and rings is important,
and it is not yet clear what exactly they mean for galaxy dynamics and
evolution (see review by Buta 2013). The least well-understood 
are the lenses, which are
morphological components with a shallow brightness gradient interior
to a sharp edge.  One formation process
suggested for the lenses in the S0s is that they formed via
bar dissolution, finally ending up to nearly axisymmetric
structures \citep{kormendy1979}.
However, this cannot be the whole story, because more than one
type of lenses (nuclear, inner and outer) exist, and often multiple lenses are found in the
same galaxies \citep{kormendy1979,lauri2011}. 
Alternatively, lenses can form via disk
instability in a similar manner as bars \citep{atha1983}, 
or via truncated star formation in the distant past \citep{bosma1983}.
However, none of the above
processes are capable of explaining multiple lenses in S0s, of which a pictorial 
example is NGC 1411 \citep{lauri2006,buta2007}. 
Recently a new lens-type, a {\it barlens}, was recognized by
\citet{lauri2011},
identified as a distinct rounder component
in the inner part of a bar. If barlenses form part of the
bar, this opens a new angle for understanding bars and lenses
in galaxy evolution, which
will be discussed in this study.

In order to study possible manifestations of secular evolution in
galaxy morphology, statistics of the structural components are
compiled, and the dimensions of the structures are studied for a
sample of $\sim$ 200 early-type disk galaxies, mainly S0s. We find
that bars are important drivers of secular evolution in galaxies. In
particular, a new explanation for the formation of inner lenses in the
non-barred galaxies is suggested.

\section{The database}

As a database we use the Near-IR S0 galaxy Survey (Laurikainen
et al. 2011; hereafter NIRS0S Atlas), which is a magnitude
(m$_B$$\leq$12.5 mag) and inclination (less than 65$^o$) limited
sample of $\sim$ 200 early-type disk galaxies, including 160 S0+S0/a
galaxies, 33 Sa spirals, and 12 late-type ellipticals\footnote{These numbers include 
29 galaxies with 12.5$<$m$_B$$<$12.7, which strictly speaking do not
belong to the magnitude-limited sample. However, none of the results
reported below would change if they are omitted from the
statistics.}. The $K_s$-band
images typically reach a surface brightnesses of 23.5 mag
arcsec$^{-2}$ in azimuthally averaged profiles, being 2-3 magnitudes
deeper than the 2 Micron All Sky Survey (2MASS\footnote{2MASS is a
  joint project of the University of Massachusetts and the Infrared
  Processing and Analysis Center/California Institute of Technology,
  funded by the National Aeronautics and Space Administration and the
  National Science Foundation.}) images.  In this study we use the new
morphological classifications of the NIRS0S Atlas, where also
the measurements of the dimensions of the structures are given.

The morphological classification is based on the de Vaucouleurs' revised
Hubble-Sandage system, but is more detailed than that.  It includes
the stage (S0$^-$ , S0$^o$ , S0$^+$ , Sa), family (SA, SAB, SB),
variety (r, rs, s), outer-ring or pseudo-ring (R, R'), possible
spindle shape, and the presence of peculiarity.  Lenses (nuclear,
inner, outer) are systematically coded, in a similar manner as in
\citet{kormendy1979} and in \citet{buta2010}.
The classification distinguishes ansae and x-shaped bars from
regularly shaped bar morphologies. Also, due to the sub-arcsecond
pixel resolution, it was possible to classify central structures like
nuclear bars, rings and lenses in a systematic manner.  The NIRS0S
Atlas includes also identification of weak bars in the residual
images, after subtracting the bulge model, obtained from the
structural decompositions. Alternatively, weak structures were
detected in unsharp mask images (a smoothed image subtracted from the
original image). These faint structures, together with the
identification of an exponential outer surface brightness profiles
form part of the photometric classification given in the NIRS0S Atlas,
which classification is generally used in this study. Examples of the
Atlas galaxies are shown in Figure 1.

Included in the analysis are also the photometric properties
of bulges, obtained for the same galaxies by \citet{lauri2010},
based on two-dimensional multi-component structural
decompositions\footnote{The
  decompositions were performed with BDBAR code (Laurikainen et
  al. 2004, 2005), written by Salo}. In the decompositions analytical functions (S\'ersic,
Ferrers, exponential) were fitted, in addition to the bulges and disks,
also to bars (including nuclear bars), lenses and ovals. 

\subsection{Definitions of the structures in the classification}

{\it L, l, nl} 

\noindent By lenses we mean flat light distributions in the
disks, with fairly sharp outer edges. Such structures form part of
the original galaxy classification \citep{sandage1961,sandage1994},
though no coding for the lenses was used at that time. Nuclear, inner
and outer lenses are denoted by nl, l, and L, respectively. In barred
galaxies the inner lenses have sizes roughly similar to bars, whereas
in non-barred galaxies the relative size compared to the galaxy
size defines the subtype. Outer lenses are clearly larger than the
inner lenses, and nuclear lenses are small central structures 
with similar sizes as the nuclear rings.

\noindent {\it bl}
 
\noindent Barlenses are lens-like structures embedded in the bars. In distinction 
to bulges they have much flatter light distributions, whereas in distinction 
to nuclear lenses they are much larger, typically covering 
50$\%$ of the bar size. Barlenses can be aligned with the main bar, or
be slightly twisted from the major axis orientation of the bar.

\noindent {\it R, r, nr} 

\noindent Rings are structures in which both the inner and outer
regions have well defined edges. The rings can be full rings or
non-complete pseudorings, of which both can further have their
subtypes (see Buta 1995).  Sometimes an inner ring is seen mainly
because of dark space around the sides of the bar.

\noindent {\it RL, rl, nrl} 

\noindent Ringlenses are intermediate types between rings and lenses.
They have no well defined inner edges, but the surface brightness
increases towards the outer regions, until the light distribution
drops, in a similar manner as in rings.  Ringlenses are divided into nuclear,
inner and outer structures (nrl, rl, RL) in a similar manner as
rings. Ringlenses have also similar subtypes as rings.

\subsection{Size measurements of the structures}

In the NIRS0S Atlas the sizes of the structures are given in the sky plane
(Tables 5 and 6 in the Atlas). The classified features were first
inspected visually in the images. If a feature was a clear ring or
ringlens, the cursor was used with IRAF routine TVMARK to outline the
ridge-line. If the feature was a lens the outer edge was mapped
instead. After that an ellipse-fitting program was used to fit the
points, which gave the center coordinates, the position angle of the
major axis, the major axis radius, and the axis ratio of the
component. For bars the edge of the bar
defined its size.

In this study the size measurements given in the NIRS0S Atlas were
converted to the plane of the galactic disk using the orientation
parameters obtained from the apparent major axis position angle and
the minor-to-major axis ratio, based on fitting ellipses to the outer
isophotes.  For the orientation parameters our deep optical (B,V) images
were used when available, otherwise the $K_s$-band NIRS0S images were
used.

\section{Frequency of the structure components} 

Fractions of bars, lenses and rings are calculated as a function of
Hubble type, and are given also in different bulge-to-total ($B/T$)
flux-ratio and in absolute magnitude bins in $K$-band. Tables 1-3 give
the values for all galaxies in NIRS0S\footnote{including the 29
  galaxies not forming part of the magnitude limited sample}, whereas in
tables 4-6 the values for the barred galaxies are given. In Tables 1 and 4 the
Hubble type bin $-$4$\leq$T$<$$-$3.5 is not shown,
because these galaxies very seldom have any sub-structures.  In
the lowest galaxy luminosity bin (Tables 3 and 6) the galaxy numbers
are shown, but no percentages are given.  The percentages in the
tables are calculated with respect to the total number of galaxies in each
bin. The fractions in Tables 1-6 were calculated also including only the S0-S0/a
galaxies (thus excluding Sa galaxies), but are not shown in the tables,
as there were no significant changes. The uncertainties are
estimated from $\Delta p = \sqrt{(1-p)p/N}$, where $p$ denotes the
fraction among $N$ galaxies. In the tables 'res' means that included
in the analysis are also the weak structures visible only in the
residual or unsharp mask images. In all the following the sub-types
r'l or R'L are included in the categories rl and RL, respectively.

The absolute magnitudes were calculated using the extended $K$-band
apparent magnitudes taken from 2MASS (Skrutskie et al. 2006). The distances are from
the Catalog of Nearby Galaxies by \citet{tully1988}, 
which uses H$_{0}$ = 75 km
s$^{-1}$ Mpc$^{-1}$.  A correction for Galactic extinction was made using
the values given in NED, based on the maps of \citet{schlegel1998}.
As explained in \citet{lauri2010}, 
the $B/T$ values are corrected for Galactic and internal extinction
using different corrections for the bulge and the disk \citep{graham2008}.
By bulge we mean the flux fitted by the S\'ersic
function, and by the disk all the remaining flux, which includes, not only
the exponential disk, but also the bars, ovals and
lenses. 

\subsection{Bars} 
     
The bar fraction (SB+SAB) we find among the S0$^+$ galaxies (T=$-$1) is as
high as for spirals (T=1), e.g. $\sim$ 80$\%$ have bars. It
gradually drops towards the earlier type systems, being 50$\pm$8$\%$
among the S0$^o$s (T=$-$2), and 35$\pm$7$\%$ among the S0$^-$ types
(T=$-$3) (Table 1). The tendency is the same separately for strong (B) and
weak (AB) bars (see Fig. 2a). The tendency of lower bar
fractions among the S0 galaxies, compared to spirals, has been
previously shown by Laurikainen et al. (2009) using near-IR images for
360 galaxies at z$\sim$0, and by Nair $\&$ Abraham (2010a; using 14000
optical images) at z=0.01-0.1. The obtained bar fractions among the 
S0s and spirals were 46$\%$ vs $\sim$65$\%$ at z=0, and 12$\%$ vs
22$\%$ at z=0.01-0.1, respectively.  In both studies the
classification and the identification of bars were done visually. The
large differences in the bar fractions
are mainly because Nair $\&$ Abraham detected only strong bars, whereas
we include also weak bars into the statistics. A fairly low bar fraction
among the S0s has been found also by \citet{aguerri2009}
at z=0.01-0.04, the fractions being 29$\%$ and 55$\%$ for the S0s and
spirals, respectively. However, in their study the Hubble type was
based on the concentration parameter (see Conselice et al. 2000).
Taking into account the large range of $B/T$flux-ratio within
one Hubble type (see Laurikainen et al. 2010) the bar fractions
obtained by Aguerri et al. and Laurikainen et al. are not directly
comparable: in Aguerri et al. the implicit association of high $B/T$
for the S0s causes a bias towards the early-type S0s, which have 
higher $B/T$ flux-ratios and a lower fraction of bars.

The bar fraction depends on the $B/T$ flux-ratio and the 
luminosity of the galaxy (Tables 2 and 3, respectively). It is clearly higher in
fainter galaxies, the fractions being 73$\pm$5$\%$ 
and 39$\pm$5$\%$ in the galaxy
luminosity bins of $-$22$>$M$_K$$\geq$$-$24 and
$-$24$>$M$_K$$\geq$$-$26, respectively. This is the case also separately
for strong (B) and weak (AB) bars, and when only the S0-S0/a
galaxies are considered. The fraction of strong bars increases
towards a lower $B/T$ flux-ratio (Fig. 2b). The S\'ersic index $n$
(Fig. 2c) is also interesting, since the bulges in the
strongly barred galaxies are more exponential than those in weakly
barred or in non-barred galaxies. Although all the family classes cover
a similar wide range of $n$-values, the peak values are shifted with respect to each
other: $n_{peak}$$\sim$1.75, 2.25, and 2.75 for B, AB, and A
families, respectively. The KS-test shows that the distributions of SA and 
SB galaxies in Figures 2b and 2c are different (p=0.1$\%$ and 3.4$\%$, respectively). 

Compared to non-barred galaxies, lower $B/T$ flux-ratios in barred early-type spirals have
been previously found by \citet{lauri2007} and \citet{weinzirl2009}.
A similar result was obtained also by \citet{coelho2011}
for disk galaxies, without specifying the morphological type. 
For the S0s somewhat different results have been obtained depending on the
exact sample used \citep{lauri2007,lauri2010}.
Based on the results of the current study, this can be understood, because the relative
fraction of strong and weak bars varies in these S0 galaxy samples:
including a large fraction of weak bars any difference in $B/T$ flux-ratio between  
barred and non-barred galaxies is easily diluted (like in Laurikainen et al. 2010).   
Bar fractions 
in the lower luminosity spirals, e.g. in very late-type spirals, have
been studied by \citet{barazza2008}: 
they found that the
galaxies with smaller $B/T$ flux-ratios have a larger fraction of
bars. It is worth noting that the bars in the lower luminosity
galaxies are not necessarily similar to those in the bright
galaxies.


\subsection{Barlenses} 

In the NIRS0S Atlas barlenses were identified as lens-like structures
inside the bar, and they were suggested to form part of the bar itself.  We
found that the occurrence of a barlens does not depend on the Hubble
type (Fig. 3a), or on the galaxy luminosity. The fractions of
barlenses in the two absolute magnitude bins are 34$\pm$6$\%$ and 25$\pm$6$\%$,
respectively (Table 6). In Table 7 the fractions of the 
structure components in barred S0-S0/a galaxies with barlenses are compared with
those without a barlens.  The clearest difference is that even
52$\pm$9$\%$ of the barlens galaxies have ansae in the main bars,
whereas among the galaxies without any barlens only 24$\pm$6$\%$ of bars have ansae.
Another noticeable characteristic is that in the barlens galaxies
multiple lenses are rare: only 24$\pm$8$\%$ of them have multiple
lenses, compared to 56$\pm$7$\%$ in barred galaxies without a barlens.  The
occurrence of a barlens is strongly connected to the properties of
the bulges so that their fraction increases towards
the lower $B/T$ flux-ratios (Fig. 3b), and lower values of the S\'ersic
index
(Fig. 3c). These characteristics are partly related to the fact that
barlenses are more common among the strongly barred galaxies
(62$\pm$9$\%$ and 38$\pm$9$\%$ among the B and AB families, respectively; see Table 9),
which also have fairly low $B/T$ flux-ratios and small values of the S\'ersic index.

\subsection{Rings, ringlenses and lenses}

\subsubsection{Frequency as a function of Hubble type }

Lenses are common features in the early-type disk galaxies (see also
Laurikainen et al. 2005, 2009, 2011; Nair $\&$ Abraham 2010b), and are
crucial for identifying the S0s.  Lenses are found to be distributed over all
Hubble types in NIRS0S, but the fully developed lenses (L, l) are more
frequent among the early-type S0s (T=$-$3). For barred galaxies this
is shown in Figure 4a (Table 4): among the Hubble type T=$-$3, even
59$\pm$12$\%$ of the barred S0s have an outer lens, and 41$\pm$12$\%$ have
an inner lens.  On the other hand, the fraction of nuclear lenses ($\sim$25$\%$) is
nearly constant as a function of Hubble type, until it drops at
T=$-$3.

A significant fraction of the inner rings and ringlenses appear at 
Hubble type T=$-$1, after which the fraction gradually drops towards the later types
(Fig. 4b,c). Outer rings also appear in the S0s, but they are even more common
in the early-type spirals.
Rings are generally thought to be resonance structures of
bars \citep{atha1985,buta1995},
or linked to resonances via 
bar-driven manifolds \citep{atha2010}.
They indeed appear mainly in barred galaxies, 
the fractions of outer, inner and nuclear rings being 14$\%$, 17$\%$ and 12$\%$,
respectively, in contrast to the ring-fractions of 1-7$\%$ among the
non-barred galaxies (Table 8).
Using optical images the ring fractions might be even higher
because many rings are known to be active star forming regions (see
Buta $\&$ Combes 1996; Grouchy et al. 2010; Comer\'on et al. 2010).
It is worth noting that although nuclear lenses appear in all Hubble types,
nuclear rings are more concentrated in late-type S0s (T=$-$1), in a similar
manner as inner rings.

\subsubsection{The parent galaxy properties}

As rings and ring-lenses are assumed to be related phenomena, also
their parent galaxy properties are expected to be similar.  Indeed, we
find that the distributions of the $B/T$ flux-ratios are similar
(Fig. 4e,f). However, the galaxies with full lenses have higher
$B/T$-flux ratios, being typically larger than 0.3 (Fig. 4d).  Again,
the tendencies are similar among the inner and outer rings, and among
the inner and outer ringlenses. In the highest $B/T$-bin there are
more inner (l) than outer lenses (L), which might be partly because
the shallow outer lenses are more difficult to detect.

The appearance of rings, ringlenses and lenses is not very sensitive
to galaxy luminosity. However, while lumping together the outer lenses
(L) and ringlenses (RL) they appear more frequently among the fainter
galaxies (32$\pm$5$\%$ vs. 19$\pm$4$\%$ in the two galaxy luminosity
bins, respectively; Table 3, ). Also, the inner rings (r,rs) appear more
frequently among the fainter galaxies (20$\pm$4$\%$ vs. 3$\pm$2$\%$, in
the two bins, respectively).  The tendencies are similar if only the
barred galaxies are considered (Table 6).

In conclusion, rings and ringlenses are
common in the early-type spirals and in the late-type S0s, whereas
the fully developed lenses are more common in the early-type
S0s. There is no strong dependence of the type of structure on 
galaxy luminosity.  However, the galaxies with full lenses typically
have higher $B/T$ flux-ratios than the galaxies with rings or
ringlenses. On the other hand, the occurrence of 
nuclear lenses does not depend on the Hubble type 
or on the properties of the bulge.

\subsection{Multi-component structures} 

\subsubsection{Multiple bars}

Multiple bars have been previously detected in $\sim$30$\%$ of the barred
galaxies and in $\sim$20$\%$ of all early-type disk galaxies, 
by \citet{erwin2002} and by \citet{laine2002}.
The sample of Erwin $\&$
Sparke consists of 30 barred S0-S0/a galaxies, of which 10 have double
bars. In order to detect also the weak bars they used color images
and applied the unsharp mask techniques. The double bar catalog by \citet{erwin2004}
is larger, including 38 multiple barred S0+S0/a
galaxies collected from the literature. However, as noticed by Erwin
the catalog is most probably biased towards peculiar and strongly
barred galaxies, since it is randomly collected from different sources.
These results are discussed also in the recent
review by \citet{erwin2011}.
The fractions of double bars in the early-type disk galaxies have been 
studied also by \citet{lauri2009}
for a sub-sample of 127 galaxies
in NIRS0S largely confirming the previous results.

Among the S0-S0/a galaxies we detect
35 multiple bars, which makes 20$\pm$3$\%$ of all, and 30$\pm$5$\%$ of
the barred S0+S0/a galaxies, fully confirming the previous
results.  However, it is worth noting that, compared to 
\citet{erwin2002},
the total number of double bars in NIRS0S is $\sim$ three times larger. 
The NIRS0S double bar sample is 
comparable in size with that of the double bar catalog by \citet{erwin2004},
but has the advantage of being magnitude-limited and therefore more
complete in a statistical sense.  
Both in the present study and in \citet{erwin2002}
unsharp masks were used to identify the faint nuclear bars.
Alternatively, we applied also a
decomposition approach for detecting the weak bars in the residual images,
after subtracting the bulge model from the original image. These
methods of detecting multiple bars are important, because even 40$\%$
of all nuclear bars are weak or overshadowed by massive
bulges in the direct images.  
\citet{erwin2002} 
also showed that the fractions of double bars are similar in S0, S0/a, and
Sa galaxies, which we confirm.  We further show that
there are no systematic differences in the multiple bar fractions between the
different sub-types of the S0s (Fig. 5a). 
In Figures 5b,c we also show that the
occurrence of a nuclear bar is not strongly connected to the properties of the bulges:
they appear in the whole range of $B/T$ flux-ratios. The S\'ersic index is peaked 
at $n$$\sim$2.25.

We detect also seven nuclear bars in galaxies without any main bar,
e.g. the bars have similar sizes as typical nuclear
bars in barred galaxies.  In Table 10 the Hubble types, the bins
of the $B/T$ flux-ratios and the galaxy luminosity bins for these galaxies are
shown. It is worth noting that 5/7 of them have very early 
Hubble types, and appear in the bin of the more luminous galaxies
in our sample.

\subsubsection{Multiple lenses}

Multiple lenses have been identified in the early-type disk
galaxies a long time ago \citep{sandage1961}
and have also been discussed in detail for some
individual galaxies by \citet{kormendy1979}.
One of the most spectacular
cases is NGC 1411, which has a series of lenses \citep{lauri2006,buta2007}.
In Figure 5 the
fractions of multiple lenses are shown, as a function of Hubble type,
and as a function of the properties of the bulges.  Barlenses are
excluded here, because they are assumed to form part of the bar.  The
following cases are shown: (1) multiple lens systems consisting of
only fully developed lenses ('multi-lens'), (2) systems in which one of the lenses 
has some ring-like characteristics ('multi-rile'), and (3)
cases in which both lenses have ring-like characteristics ('multi-rl').

We find that although multiple bars appear in all Hubble types in a
similar manner, the fully developed multiple lenses are more frequent
among the early-type S0s (T=$-$3, $-$2). However, including also
ringlenses, they are more uniformly distributed among all Hubble
types. Also, while multiple bars cover a large variety of $B/T$
flux-ratios, multiple lenses are more common in galaxies with $B/T$
$>$ 0.3. Furthermore, the S\'ersic index in multiple-lens galaxies is
peaked at larger values, $n$$\sim$2.75, as compared to $n$$\sim$2.25
for the galaxies with multiple bars. For the barred galaxies this is
shown in Figure 5 (Tables 4 and 5).

In the NIRS0S Atlas multiple lenses were found in 25$\pm$3$\%$ 
of the S0-S0/a galaxies (the above categories 1 and 2 combined).
Focusing only on fully developed lenses (category 1) the fraction
is reduced to 11$\pm$3$\%$. 
Even 61$\%$ of all multiple lenses (all 3 above categories combined) appear in barred galaxies. 
Of the fully developed lenses the fraction is somewhat lower, 42$\%$ of them
being barred.    

\subsection{Structures in the A, AB and B families}

An important new result of this study is that nuclear bars appear
preferentially in galaxies with weak main bars. Namely, we found that
even 44$\pm$8$\%$ 
of the galaxies with weak bars (AB), and  only 24$\pm$6$\%$
with strong (B) bars have nuclear bars (Table 9). These numbers
include the faint nuclear bars detected only in the residual images.
Notice that 
\citet{erwin2002}
found equal nuclear bar fractions among the
strongly and weakly barred systems. Possible reasons for this
difference, compared to our result, are that the sample by Erwin $\&$ Sparke is smaller (10
double barred galaxies) and might be biased towards peculiar strongly
barred galaxies that have a smaller number of multiple bars.  Also,
for one half of the galaxies optical images were used in their study and
therefore some of the weak nuclear bars might have also been hidden by
dust. Most probably the larger sample by \citet{erwin2004}
is missing galaxies with weak main bars, and therefore 
is not suitable to study the effect.

The barlens is an other structure component for which the frequencies are
different among the strongly and weakly barred galaxies: the fractions
are 62$\pm$9$\%$ vs. 38$\pm$9$\%$ among the B and AB families,
respectively.  The differences in the fractions of all the other
structures components between the B and AB families are only marginal.
It is also interesting that although the fractions of the structure
components in the non-barred galaxies (A) are typically very small
(3-10$\%$), even 21$\%$ of them have inner lenses (l). In fact, the
fraction of the inner lenses is found to be constant among all the
family classes (see Table 9).

\section{Dimensions of the structures}

The dimensions of the structure components are studied using the
measurements given in the NIRS0S Atlas (Tables 5 and 6), after converting
them to the plane of the galactic disk.  No internal extinction
corrections were applied to the lengths. In the Atlas two additional
length estimations are provided for many of the bars, based on the
radial profiles of the ellipticities, but as only the visual length
estimation is available for all bars, it is used in this study.

\subsection{Bars}

The main bars were found to have a mean size (=semi-major axis length) of $\sim$ 4 kpc, in
agreement with the previous studies (see for example Erwin 2005,
2011). No difference was found in the main bar sizes between single
and multiple bar galaxies, the mean sizes being 4.1$\pm$0.4 kpc and
4.4$\pm$0.4 kpc, respectively (see Table 11). This differs from
\citet{erwin2011}
who found smaller bars in galaxies with single bars,
with typical sizes of $\sim$2.5 kpc. A possible reason for this
discrepancy is that some single bars in the NIRS0S Atlas were
classified as nuclear bars, based on their small sizes in respect of
the galaxy diameter. Some of such small bars in the sample 
by Erwin might have been considered as main bars.

The sizes of nuclear bars in our sample cover the range of 0.1-1.5
kpc, (with the mean of 0.7$\pm$0.4 kpc), comparable with that obtained
by \citet{erwin2005,erwin2011}
who give a size range of 0.1-1.2
kpc. Nuclear bars in galaxies that do not have any main bar, have the
sizes of 0.3-0.9 kpc, well within the range of typical sizes of
nuclear bars.

\subsection{Rings and lenses, normalized to bar length}

\subsubsection{Comparison to Kormendy (1979)}

The dimensions of the inner rings (r), lenses (l) and ringlenses (rl) have
been previously studied by \citet{kormendy1979}
for a sample of 120
disk systems, including 41 S0-S0/a galaxies. These early-type systems form part of
the NIRS0S Atlas, thus allowing us to compare our measurements with those obtained by
Kormendy. In both studies the lengths were measured in a
similar manner. In cases where the assigned Hubble type differed between the two
studies, we adopted the one given in NIRS0S Atlas (the classification by
Kormendy was based on The Second Reference Catalogue of Bright Galaxies
by de Vaucouleurs, de Vaucouleurs $\&$ Corwin 1976). 
Kormendy (1979) normalized the sizes of the structures to the bar size, without
deprojecting the images to the disk plane, and
the major or minor axis length of the ring/lens
was chosen depending on which one was
closer to the bar size. This guaranteed that
the relative lens/bar size was correctly estimated regardless of the
viewing angle.

The comparison to Kormendy (1979) is shown in Figure 6a, where the length of the
structure is normalized to the bar size.  It appears that in the overlapping sub-sample
of NIRS0S the lengths
of the rings, lenses and ring-lenses are very similar in the two
studies.  The peak values are {\it length}(r,rl,l)/{\it length}(bar)
$\sim$1 for all structures, e.g.  they
typically all end up to the bar radius. This similarity in the lengths
of bars and inner lenses was used by \citet{kormendy1979}
to suggest that lenses
are formed from bars, e.g. there is a process which makes some bars to
evolve rapidly into a nearly axisymmetric state. Kormendy excluded 
three outliers, NGC 936, NGC 4262 and NGC 3412, having {\it length}/{\it
  length}(bar)$>$1.3. For the
first two galaxies the lens was considered to be an outer lens by him, and
for NGC 3412 his barlength measurement was uncertain. 
In our
measurements there is one outlier, NGC 4612, but in this galaxy the
lens is truly larger than the bar. As this galaxy has also an outer
ringlens, the structure we assign as an inner lens cannot be 
a misclassified outer lens. However, it is worth noting that 
although our length measurements are in a good agreement with
those of \citet{kormendy1979}
the structures classified as lenses (l) by Kormendy
(67$\pm$9$\%$ are lenses) are largely ringlenses (rl) or rings
(r) in our classification (only 18$\pm$8$\%$ are lenses).  
It seems that we have
interpreted the broad dispersed rings as real rings or ringlenses,
whereas Kormendy has interpreted them as lenses. 
In total the sample by Kormendy
has only a few galaxies having structures that we call as lenses.
In his study Kormendy used red and blue Palomar Sky Survey (POSS) copy plates.

\subsubsection{Using complete NIRS0S sample}

Next we look at the normalized sizes of the structures in the complete
NIRS0S. The following peak values were found: {\it
  length}(r,rl,l)/{\it length}(bar)=0.9, 1.1 and 1.3 for rl, r and l,
respectively (see Fig. 6b; our bins are 0.2 units wide), indicating
that in fact only the rings and ringlenses end at the bar radius,
whereas the fully developed lenses are clearly larger than the bar.
Using the KS-test the difference for the lenses and rings is found to
be statistically significant: the probability that their distributions
are drawn from the same population is only p=0.07$\%$. For the
distributions of ringlenses and lenses the difference is not
statistically significant (p=4.4$\%$), nor for the distributions of
rings and ringlenses (p=86$\%$). Although the size difference between
the rings and lenses might be partly due to the different ways of
measuring the structures (edge of the structure for bars and lenses,
and the ridge-line for rings and ringlenses), it does not wash out the
conclusion that the lenses are on average slightly larger than the
bar.
All structures have also small tails
towards larger normalized lengths.  Some galaxies were eliminated in
the figure for the following reasons: a) the bar is subtle, peculiar,
or has an unfavorable orientation in the sky (NGC 1537, NGC
4203, NGC 3489, NGC 6012 and NGC 6438).  b) The dimension of the lens
is uncertain (NGC 1371), and c) the lens most probably
is an outer lens (NGC 1389). A particularly high value of {\it length}(l)/{\it
 length}(bar)=1.9 was measured for NGC 5750,
but in this galaxy the lens is truly larger than the bar.
The galaxy IC 5328, which in the NIRS0S Atlas was interpreted to have a lens
({\it length}(l)/{\it length}(bar)=0.35), was excluded because the small lens is in fact 
a barlens.  One can ask whether the fairly large inner lenses 
({\it length}(l)/{\it length}(bar)$\sim$1.3-1.5) are actually 
outer lenses. The answer is that most probably not, because proper
outer lenses typically have dimensions of the outer rings, or are even
larger.

For the outer rings the normalized lengths peak at {\it length}(R)/{\it length}(bar)=2.2, which is
in a good agreement with the
mean value of 2.21$\pm$0.12 obtained by \citet{kormendy1979}
and Buta (1995) in the B-band.
For the outer ringlenses (RL) and outer lenses
(L) the variations in the lengths are large, the distributions lacking any clear peaks.
In those non-barred
galaxies in which a series of lenses are detected, the terminology of
inner and outer lenses is merely academic.

\subsection{Rings and lenses, normalized to galaxy size}

In order to compare the dimensions of rings and lenses in barred and
non-barred galaxies, the lengths were normalized to the galaxy size,
R$_{25}$. This is the galaxy radius at the surface brightness of 25 
magnitudes arcsec$^{-2}$ in the $B$-band, taken from the Third Reference Cataloque of
Bright Galaxies (de Vaucouleurs et al. 1991). In barred
galaxies the bar defines the lens type: the inner lens is located at,
or near to the bar radius, whereas the outer lens has typically twice
the size of the bar, in a similar manner as the inner and outer rings 
\citep{kormendy1979,buta1995,lauri2011}.
However, in the non-barred galaxies the smaller lens is
simply called the inner lens and the larger one is the outer lens.  If only
one lens is present, its size relative to the galaxy size defines
the inner/outer characteristic.  An alternative would be to use some
limit in kiloparsecs, based on a typical bar
size in galaxies. However, taking into account that the relative bar
size varies a lot from one galaxy to an other, this is not necessarily a
better approach.

The number histograms of the lens dimensions in the barred and
non-barred galaxies are compared in Figure 7 where, for better
statistics, lenses and ringlenses are grouped together. The comparison
is made first by lumping together the inner
and outer structures, which shows that the lenses in the
non-barred galaxies are in general clearly smaller (Fig. 7a). This might be partly
because more outer lenses are detected in barred galaxies, but that
does not explain all of the size difference. This becomes clear when
comparing the inner structures (for which better statistics is available) 
separately for barred and non-barred
galaxies (Figs. 7b,c): while the peak value for the non-barred
galaxies is 0.2, for barred galaxies it is 0.4 (the mean values are
0.28$\pm$0.03 and 0.41$\pm$0.03, respectively).  

The blue histogram in Figure 7b shows the distribution of barlenses
in barred galaxies (for comparison the same histogram is over-plotted
on Fig. 7c for non-barred galaxies). By definition,
r$_{barlens}$ $<<$ r$_{l}$ is expected.  The interesting thing is that
the length distribution of the inner lenses in the
{\it non-barred galaxies} (Fig. 7c) is very similar to that for barlenses 
in {\it barred galaxies} (the mean
value is 0.20$\pm$0.01). This is a puzzle that needs to be explained.

\subsection{Relative dimensions of lenses in multiple systems}

In the following the relative dimensions of lenses in multiple systems
are compared in barred and non-barred galaxies.  We are particularly
interested in the inner and outer lenses, excluding the nuclear
lenses, which leads to the total number of 17 galaxies. This number
includes the cases in which the multiple systems are
a combination of full (l,L) and intermediate type
(rl,RL) lenses. There is a tendency showing that the mean ratio of
the outer to inner lenses is larger for the non-barred galaxies,
e.g., $<${\it length}(L,RL)/{\it length}(l,rl)$>$ = 2.5$\pm$0.2 (N=8),
as compared to 1.9$\pm$0.2 (N=8) for the barred galaxies (see Table
11). We can also look at the length ratio separately for those multi-lens systems,
in which both lenses are fully developed, which gives $<${\it
  length}(L)/{\it length}(l)$>$=2.85$\pm$0.19 (N=4).
Taking into account the small number statistics the dispersions
(what is given is the standard deviation of the mean, equal to
STD/$\sqrt{N}$) are fairly small. For NGC 1574
the length ratio is $\sim$5. This galaxy was excluded because the dimension of
the outer lens was hard to define accurately.  

The average length ratio 1.9 obtained for multiple lenses in barred galaxies 
is similar to that obtained by us 
for the outer to inner rings, e.g. {\it length}(R)/{\it
  length}(r)$\sim$2.0 (calculated from the peak values of the length
distributions for the inner and outer rings, both normalized to
barlength). 
This value is in
good agreement with the previous measurements by
\citet{kormendy1979}, \citet{atha1982}, and \citet{buta1995} for the rings.

\section{Galaxy luminosities and the sizes of the structures}

\subsection{Galaxy luminosity distributions for barred and non-barred galaxies}

Distributions of the galaxy luminosities for the barred and non-barred
galaxies in the $K$-band are shown in Figure 8. Clearly, there is no
upper limit in galaxy luminosity (mass) for bar formation. In  
\citet{mendez2010}
such a cutoff limit was reported to appear at M$_r$=$-$22 mag,
corresponding to M$_K$=$-$25 mag. 
Also, barred and non-barred galaxies cover nearly the same range of total galaxy
luminosities. However, there is a tendency of increasing mean galaxy
luminosity from strongly barred (B) towards weakly barred
(AB), and non-barred (A) galaxies. The mean absolute magnitudes in the $K$-band are
(the uncertainties are standard deviations of the mean):

$<$M$_K$$>$ (A) = $-$24.11$\pm$0.15 

$<$M$_K$$>$ (AB) = $-$23.78$\pm$0.12 

$<$M$_K$$>$ (B) = $-$23.52$\pm$0.14. 

\noindent It thus appears that the non-barred galaxies are on average 0.6
mag brighter than the strongly barred systems. 
For the NIRS0S Atlas
the luminosity difference between the barred and non-barred galaxies 
was indicated also by \citet{vandenbergh2012}.
The difference in the mean galaxy luminosities between the barred (B+AB)
and non-barred (A) S0s was shown also by \citet{barway2011}.

As for the main bars, also for the nuclear bars there is no cutoff
magnitude limit for the bar formation (Fig. 8). Also, the galaxy luminosity 
distributions are similar for the barred galaxies in general, and 
separately for the galaxies with nuclear bars. 

\subsection{Galaxy luminosities and sizes of the structure components}

It has been suggested by \citet{kormendy1979}
that galaxy mass uniquely determines the bar size, and also the sizes of
all the other components associated with the bar, e.g., the inner and
outer lenses, and the inner and outer rings. He tested this idea 
showing a correlation between the $B$-band
galaxy brightness and the length of the structure.  The correlations
for all the structure components were found to be tight within one Hubble
type bin. Later type galaxies were
systematically shifted towards larger galaxy brightnesses, which
was interpreted to follow from their disks
having a larger amount of recent star formation. In this study we can
re-investigate this issue in the near-IR, which is much less affected by star
formation.  We also extend the study to non-barred
galaxies and to different sub-types of the S0s.

\subsubsection{Bars}

We confirm the previous result of \citet{kormendy1979}
in a sense that a
clear correlation exists between the absolute galaxy magnitude and the length
of the bar (Fig. 9). Overall the S0s occupy roughly the same
region as the early-type spirals (T=1). However, for a given
barlength, the early-type S0s (T=$-$3, $-$2) are brighter than the
late-type S0s (T=$-$1), opposite to the trend found by Kormendy in the
optical region.  For spirals and for S0 galaxy sub-types the following
relations are found:

    T=1: M$_K$ = $-$4.0 log D - 21.6, 

    T=$-$1: M$_K$ = $-$9.1 log D - 18.0, 

    T=$-$3: M$_K$ = $-$3.4 log D - 22.9 



\noindent where D stands for bar length in kpc.
Combining T=$-$3, $-$2 the coefficients of correlation are $-$3.6 and $-$22.5, 
and for T=$-$1,0 they are $-$4.6 and $-$21.6.   
For each morphological type the
slope is different from that obtained by \citet{holmberg1975}
between the galaxy luminosity and its diameter, implying that the
surface brightness of the bar is not constant. Compared to the Holmberg 
relation (dN/dlogD=$-$5) the correlation is shallower for T=$-$1 and steeper for T=$-$3.
It is worth noting
that although the correlation for T=$-$1 is tight, it does not fall 
between the regression line of the
earlier and later Hubble types. This might be related to the
fact that the bar length (normalized to galaxy size) maximum appears at
T=$-$1 (see Fig. 5 in Laurikainen et al. 2007), e.g. all bars in these 
Hubble types are long, independent of the galaxy luminosity.
But still, the physical meaning of this behavior is unclear.

\subsubsection{Rings and lenses}

As for bars, also for the inner rings, ringlenses, and lenses correlations exist
between the size of the structure and the galaxy luminosity.  The
dispersions are large, but they are reduced while plotting the different
Hubble types separately (Fig. 10a). There is no obvious difference in
the behavior of rings (r), ringlenses (rl) and lenses (l)
(Fig. 10b). However, while comparing the sizes 
among the barred and non-barred galaxies (Fig. 10c)
it appears that the inner structures are smaller for the brightest
non-barred galaxies.  Similar tendencies as for the inner structures
were found also for the outer rings (R), ringlenses (RL) and lenses
(L) (Fig. 10d,e,f).

We conclude that although the sizes of bars, rings, and lenses depend
on the galaxy luminosity, the luminosity does not uniquely define their
sizes.  In barred galaxies the size of the structure depends also on
the Hubble type (smaller in the early-type S0s), and in non-barred
galaxies the lenses are smaller in the most luminous galaxies.

\section{Discussion}

Secular galaxy evolution is actively debated both from the
observational and the theoretical points of view. Large galaxy surveys
are under progress, focusing on this topic via galaxy morphology at
different redshifts, largely based on automatic analysis approaches.
However, it is also important to study the nearby galaxies, for which
more detailed analysis can be made. Bars are expected to be important
driving forces of secular evolution. They are efficient drivers of
gas to the central regions of the galaxies \citep{shlosman1989,shlosman1990},
possibly adding mass to the bulge, or even creating nuclear
bars \citep{friedli1993}.
Bars can drive spiral arms (Salo et al. 2010 and references there), or they 
can be in the dynamical interaction with the halos 
\citep{atha2003,inma2006} as well as with the bulges
\citep{atha2002,atha2003,saha2012},
leading to re-distribution of stellar matter in galaxies. In the following
possible manifestations of secular evolution in the morphology of the
S0 galaxies are discussed.

\subsection{Do bars induce bulge growth?}

The fact that the bulges in S0s are on average fairly 
exponential ($<$n$>$$\sim$ 2)
is consistent with the idea that many of them are (or contain) pseudobulges 
(or in terminology of Athanassoula 2005, disky bulges) triggered by bars. However,
the result that the $B/T$ flux-ratio is smaller in the strongly
barred S0s (compared to weakly barred or non-barred S0s), is seemingly 
in contrast with bar induced bulge growth, because 
strong bars are expected to be more efficient drivers of gas to the
central regions of the galaxies \citep{shlosman1989,shlosman1990}.

Nevertheless, there is an alternative point of view. Galaxies that
have relatively small (non-classical) bulges have slowly rising
rotation curves, provided that their dark matter halo is not too 
centrally concentrated or cuspy). In that case no Inner Lindblad Resonance (ILR)
barrier is expected for inward moving trailing perturbation wave
packets, which can thus reflect from the galaxy center as fresh
leading packets (Toomre 1981; Salo $\&$ Laurikainen 2000; see also review
by Athanassoula 1984). If the disk is reactive (low value
of Toomre $Q$ parameter) this may lead to an efficient swing
amplification cycle \citep{toomre1981}, and a strong secular bar growth
(Athanassoula 2012). Because of its strength, such a bar will
eventually lead to a growth of a relatively massive pseudo-bulge,
but since the galaxy will have no classical bulge its $B/T$ flux-ratio
will remain small. On the other hand, galaxies which
initially have a massive (or classical) bulge have a more steeply
rising rotation curves, implying an ILR for all small to intermediate
large pattern speeds, likely to inhibit bar formation or allow only
weak bars to grow.  Therefore it is possible that the bulges we observe
in strongly barred galaxies were induced by bars, whereas the bulges in
weakly barred systems were largely formed at the epoch prior to bar
formation.  In this context the recent observations by
\citet{perez2011}
are interesting, because they
have shown that the bulges in barred early-type galaxies are more
metal rich and more $\alpha$ enhanced than the bulges of their non-barred
counterparts. This indicates that the bulges of typical barred
galaxies were formed later than the bulges of non-barred galaxies,
possibly in a rapid starburst.

Disk-like pseudobulges form also in cosmological simulations due to
massive starbursts at high redshifts, without invoking any secular
evolution \citep{okamoto2012}.
These bulges can be maintained up to
the redshift zero, in case that the galaxies have not had any
subsequent merger events with mass ratios larger than 0.1. However,
such simulated bulges are generally more massive than the observed
bulges, both in spirals and in S0s \citep{governato2007,scanna2011}.

\subsection{Do main bars trigger nuclear bars?}

There are two main types of N-body simulations producing long-lived
nuclear bars: those in which the initial conditions of the simulations
include a specific inner component, be it a disk or a pseudo-bulge,
whose instability could form an inner bar \citep{friedli1993,debattista2007,shen2009}.
Or, those in which the
nuclear bar forms spontaneously from standard initial conditions
\citep {heller2001,pertti2002,englmaier2004,Heller.SA.07L,Heller.SA.07}.
It is, however, still an open question, as well as one that needs to be
observationally tested, what physical conditions in galaxies are necessary
in order to create long-lasting double bars.

Double bars appear in 20$\%$ of the disk galaxies 
(Laine et al. 2002; Erwin $\&$ Sparke; Laurikainen et al. 2009; this study), and
therefore are hard to understand merely as transient features in
galaxies, as suggested based on the early simulations by Friedli $\&$
Martinet (1993) and Englmeier $\&$ Shlosman (2004). In this study we
have additionally shown that nuclear bars appear more frequently among
the weakly barred (AB) than among the strongly (B) barred galaxies
(44$\%$ and 24$\%$, among the AB and B families, respectively). 

From the point of view of our observations the theoretical models 
by \citet{macie2008}
are the most interesting.  In
their models both the nuclear and the main bar are supported by loops,
equivalent to periodic orbits, but for the case of double rather than
single bars (the terminology of loops is from 
\citet{maciie1997}).
Maciejewski $\&$ Athanassoula found that the most stable
models are those in which the main bar is not too massive, and/or the
ellipticity of the bar is not very high (their models 16 and 17), thus
diminishing the fraction of phase-space with chaotic orbits.  This
kind of bars are very much like the AB family of bars, which in the
present study are found to have a large number of nuclear bars.

An important question is also why do we see nuclear sized bars in
galaxies without any main bar. Such galaxies in our sample are: NGC
484, NGC 1553, NGC 2902, NGC 3169, NGC 3998, NGC 4694 and
NGC 5333, including the faint bars detected only in the residual
images. 
Characteristic for these galaxies is that they are often luminous and
have quite large $B/T$ flux-ratios
($<B/T>$=0.44$\pm$0.06) and relatively large values of the S\'ersic index
($<n>$=2.6$\pm$0.2). The values are similar or even larger than those in the early-type
S0s (for T=$-$3 $<B/T>$=0.39 and $<n>$=2.2), and significantly
larger than for the late-type S0s (see Laurikainen et al. 2010: for T=$-$1 $<B/T>$=0.28, and $<n>$=2.1).
A possible reason why the nuclear sized bars 
in the 'non-barred' galaxies are so small might be related to the impact of the 
massive bulges to the rotation curves, which not only makes it harder
to create bars (as discussed in Section 6.1), but also makes 
the bars smaller. This is because of their higher 
bar pattern speeds (must exceed the maximum of $\Omega$ - $\kappa$/2),
which implies shorter co-rotation distances, and therefore also shorter bars.

\subsection{Formation of lenses in barred S0s}

In his pioneering paper \citet{kormendy1979}
suggested that
lenses in barred galaxies form when the stars in bars are gradually
spilled out, forming more axisymmetric structures surrounding the
bar. In this picture the inner lenses are expected to have similar
dimensions as the bars. The similarity of the lengths of bars
and bar-related inner structures was shown by Kormendy in the optical
region for a large range of Hubble types, including the S0s. He
also concluded that the galaxy mass (luminosity) uniquely
defines the sizes of bars and of all the structures related to bars.

However, in this study we found that the fully developed lenses are on average a
factor of 1.3 larger than the bar. We also showed that in the near-IR the
galaxy luminosity does not uniquely define the sizes of bars and the
bar related structures: the size depends also on galaxy morphology
(smaller in the early-type S0s with T=$-$3, $-$2). Therefore, our
observations do not give any clear support to the mechanism suggested
by \citet{kormendy1979}.

Lenses in barred galaxies form also in the simulation models 
by \citet{atha1983},
most likely via disk instability in a
similar manner as bars. As the main difference between bars and lenses
is in the ellipticity, in her models the lens formation can be accounted
for by a larger amount of random motions initially present in the
disk, e.g., the disk is dynamically hot.  If both cool and hot
components are present, a bar and a lens can appear in the same
galaxy, with the same orientations and lengths. Thus the factor of
1.3 difference in the observed sizes of bars and inner lenses can be a
problem also in her models.

Lenses in some barred galaxies might also be resonance-related structures, 
formed by dynamical heating from rings, which are gradually transformed 
into lenses via a ringlens phase \citep{buta1995,lauri2011}.
It is easy to find pictorial examples of such possible evolution
(see Fig. 11), but we have also other observational evidence supporting
this hypothesis. We showed that the mean ratio for the lengths of the outer
(L,RL) and inner (l,rl) lenses is $\sim$1.9, which is very
close to that obtained for the outer and inner rings (ratio$\sim$2.0) 
\citep{kormendy1979,atha1982,buta1995}.
In the linear resonance theory the predicted length ratio for the Outer
Lindblad Resonance (OLR) and the Corotation Radius (CR) is {\it
  length}(OLR)/{\it length}(CR)= 1 + $\sqrt{2}$/2  $\sim$ 1.7, for a flat rotation curve. 
According to typical slopes of
the rotation curves of Sa-Sb galaxies the predicted range is
between 1.7-2.3 \citep{atha1982}.
The fact that the rings and ringlenses appear in the same Hubble types
(see Fig. 12) is consistent with this picture. However, the fully developed
lenses are clearly more common in the early-type S0s, and have a larger 
length difference between the outer and inner components than in case of 
the rings.
This implies that not all lenses can be formed via a ringlens phase.

\subsection{Lenses in the non-barred S0s: relics of partially dissolved bars?}

In the scenario by \citet{kormendy1979}
the lenses in the non-barred galaxies are the end products of the 
transformation process, e.g.  all the stars of
the bar have been eventually spilled out into a lens.  Other suggestions for the
formation of lenses in the non-barred galaxies are those given by
\citet{atha1983} and \citet{bosma1983}.
In the models by Athanassoula it would mean that the galaxy
is dynamically too hot for creating bars, but 
still cool enough for lens formation.  \citet{bosma1983}
showed first observational evidence that lenses in the non-barred
galaxies are dynamically distinct from the outer disks. As the lenses were
redder than the disks, he suggested that the
primary components formed relatively early in the Universe 
by truncated star formation.
Lenses may also form via ringlenses or by winding of the spiral arms in the
local Universe \citep{lauri2011}.
However, none of these scenarios can readily explain the formation of multiple lenses, 
or in particular why the inner lenses in the non-barred S0s 
are as small as those found in this study. 

We find the normalized sizes of the inner lenses in the non-barred
galaxies to be surprisingly similar to the sizes of barlenses in the
barred galaxies: the sizes of both structures peak at {\it
  length}/R$_{25}$=0.2 R$_{25}$ (see Fig. 7c).  This allows to
speculate that the inner lenses in the non-barred S0s might actually
be former barlenses in galaxies in which the outer bar component has
disappeared, possibly during the bar evolution.  The typical
dimensions of these structures are illustrated in Figure 13, comparing
in the same scale a barred galaxy with a barlens (NGC 1452), and a
non-barred galaxy with an inner lens (NGC 524). This hypothetical
partial bar destruction phase could be related to a sequence of ansae morphology:
from pointy to more and more azimuthally dispersed appearance.
Although there is no theoretical explanation for this (see the review
by Athanassoula 2012), it is possible to find tentative examples of
galaxies, in which such a process might be under progress. For
example, in NGC 1079 it looks like the ansae had already started to
disperse into the surrounding inner ring (see Fig. 11). On the other
hand, there exist also barred galaxies with barlenses in which the
ansae under-fill the inner ring (NGC 7098).

NGC 524 is one of the brightest galaxies in NIRS0S, having both
nuclear, inner and outer lenses.  The normalized size of the inner
lens is 0.2 R$_{25}$. NGC 524 has been recently studied spectroscopically
by \citet{katkov2011}
covering a radial extent of $\sim$30''. Within this region, in
our morphological classification (NIRS0S Atlas)
we detect the small bulge at r$<$10'', and the inner lens at r$\sim$25''. The
nuclear lens inside the bulge has a dimension of r$\sim$6.5''. The outer
lens at r$\sim$57'' was not covered by Katkov et al. We can see that both
the bulge and the inner lens of NGC 524 have very old stellar populations ($>$15
and 14 Gyrs, respectively) and nearly solar metalicities.  Similar old
stellar populations of the bulges and disks in some other S0s have been
detected by \citet{olga2011}, and for bars in S0s and spirals by
\citet{sanchez2011} and \citet{lorenzo2012}.
The fact that the ages of the stellar populations of bars and lenses
are so similar (at least in the galaxies studied so far), 
is consistent with the idea that the inner lenses in the non-barred 
galaxies are partially destroyed bars, e.g. are former barlenses. 
However, the observed stellar populations would be consistent also with the alternative 
suggestions by Athanassoula (1983), Bosma (1983), and Kormendy (1979).

\subsection{Do S0s with different luminosities also have different formative processes?}

Such a dichotomy for the S0s was suggested by \citet{barway2007},
using the galaxy magnitude M$_K$=$-$24.5 as a dividing line. The
suggestion was based on the argument that the luminous and less
luminous galaxies have opposite trends in the scaling relation between
the scale length of the disk (h$_R$) and the effective radius of the
bulge (r$_{eff}$). However, in NIRS0S, using a few magnitudes deeper
images and a much larger sample of S0s, such a dichotomy was not found
\citep{lauri2010}.  It is also worth noting that in Barway et al.
the galaxies in both luminosity bins have S\'ersic indexes larger than
3.2, indicating that the bulges in none of the two groups are
pseudobulges in terms of having small S\'ersic indexes.  In Barway et
al. 2-component bulge-disk decompositions were made, while we used a
multi-component approach, allowing also fitting of bars and lenses.

However, the galaxy luminosity difference of 0.6 mag between the
strongly barred and non-barred galaxies observed in this study (see
also van den Bergh 2012, and Barway et al. 2009) is real.  In van den
Bergh (2012, calculated for NIRS0S galaxies) it was partly attributed
to large uncertainties in galaxy classifications between Sandage, de
Vaucouleurs, and us. However, a comparison of NIRS0S classifications
with those in a Revised Shapley Ames catalog of bright galaxies (RSA) 
and RC3 shows that the Hubble stage in NIRS0S 
generally agrees with that given by Sandage, because in both studies
special attention was payed to identification of lenses (which were
largely ignored by de Vaucouleurs). On the other hand, bar
classifications in the NIRS0S Atlas are more consistent with those given in RC3:
the galaxies classified as AB by us and by de Vaucouleurs are
typically non-barred in RSA (the weak AB bar category is lacking in
RSA). Therefore we don't see any major classification problem in
NIRS0S.

It is clear that the bright non-barred S0s cannot be merely stripped
spirals in which the star formation has been ceased. Aguerri et al. (2001) and
\citet{eliche2012} showed that dry intermediate and minor mergers can
induce bulge growth so that a galaxy with S0c morphological type can
develop into S0b, and S0b into S0a, without significantly increasing
the S\'ersic index of the bulge. This is consistent with the observed
$B/T$ flux-ratios in S0s, including some very small values, forming
part of the parallel sequence of S0s and spirals (Laurikainen et
al. 2010; Kormendy $\&$ Bender 2012). However, bulges in the S0$^-$
galaxies ($<$B/T$>$$\sim$0.39) are on average more massive than in the early-type
spirals which is difficult to account for by minor mergers alone, particularly
because the galaxies in NIRS0S appear mostly outside dense galaxy
clusters.  On the other hand, if these galaxies were formed by major
mergers, the fairly small S\'ersic indexes ($<$n$>$$\sim$2.2, see
Laurikainen et al. 2010), and the large number of lenses in the very
early-type S0s (among T=$-$3 even 30$\%$ have inner and outer lenses)
would be difficult to explain.  Alternatively, the most luminous
non-barred galaxies were formed first, and also evolved more
rapidly \citep{cowie1996}. 

It is worth noticing that in the current study fairly bright galaxies
have been discussed.  It is still possible that the lower luminosity
S0s, near the borderline to the dwarf early-type galaxies (dEs), were
formed in a different manner (see Barazza et al. 2008; Kannappan et
al. 2009). The structures of dEs, based on multi-component
decompositions, has been recently studied by Janz et al. (2012).

\section{Summary and conclusions}

The fractions of structural components and their
dimensions are studied using NIRS0S, which is a magnitude
(m$_B$$\leq$12.5 mag) and inclination limited (less than 65$^0$)
sample of $\sim$200 early-type disk galaxies, including 160 S0+S0/a
galaxies. We use the morphological classifications and
the measurements of the dimensions of bars, rings, ringlenses and lenses,
given in the NIRS0S Atlas \citep{lauri2011}, after first converting them to the
plane of the galactic disk.
The dust corrected properties of the bulges are taken from \citet{lauri2010},
based on two-dimensional multi-component structural
decompositions. To our knowledge this is the first statistical study
of the multiple structure components of the S0s particularly focusing on lenses.

 \vskip 0.35cm
Our main conclusions are the following:
 \vskip 0.4cm

{\it (1) Lenses in barred galaxies:}

\noindent Inner lenses (l) in barred S0s are found to be on average a
factor of 1.3 larger than the bars (Fig. 6b). This is not consistent with
the formative processes of lenses suggested by Kormendy (1979)
and Athanassoula (1983), which favor similar
dimensions for bars and lenses.
On the other hand, inner rings (r) and ringlenses (rl) have similar sizes as
bars. In barred multiple-lens systems we find observational evidence of the resonant
nature of the lenses: {\it length}(L,RL)/{\it length}(l,rl)=1.9$\pm$0.2, 
which is similar to the outer and inner rings ( {\it length}(R)/{\it length}(r)$\sim$2), 
generally associated to known resonances of the rotating bar, or linked to resonances via 
bar-driven manifolds.
 \vskip 0.4cm

{\it (2) Inner lenses in the non-barred S0s: 
barlenses in former barred galaxies, in which the outer bar component has been destroyed? }

\noindent The normalized sizes of the inner lenses (l) in the
non-barred galaxies are found to have similar sizes as barlenses (bl)
in barred galaxies, peaked at {\it length}/R$_{25}$=0.2 R$_{25}$.  As an
example we discuss NGC 524, which is one of the brightest galaxies in
our sample, and in which galaxy the lens contains similar old stellar
population \citep{katkov2011} as bars in some observed S0s and early- and
intermediate type spirals.
 \vskip 0.4cm

{\it (3) Inner lenses in the family classes:} the fractions are found
to be nearly constant, being 21$\pm$4$\%$, 21$\pm$5$\%$ and
16$\pm$4$\%$ in A, AB and B families, respectively. In the non-barred
galaxies the fraction of inner lenses is enhanced in respect of all
the other structure components.

 \vskip 0.4cm
{\it (4) Main bars:}

\noindent The bar fraction gradually drops from the Sa-S0/a galaxies (80$\%$ have bars)
towards the S0$^-$ types (35$\%$ have bars). The $B/T$ flux-ratio is smaller in the 
strongly barred (B) than in the weakly barred (AB) or non-barred (A)
S0s. Bulges in the strongly barred S0s are interpreted to be
triggered by bars, whereas in the weakly barred systems they are
suggested to have formed in cumulative accretion events prior to the bar formation.
\vskip 0.25cm

\noindent The mean semi-major axis length of the main bars is
$\sim$4 kpc, with no difference in size between the single and
double-barred galaxies (4.1$\pm$0.4 kpc vs. 4.4$\pm$0.4 kpc,
respectively). The mean size of the nuclear bars is 0.7$\pm$0.4 kpc
(0.1-1.5 kpc). The sizes of nuclear bars in non-barred galaxies were 
found to cover the range 0.3-0.9 kpc. 

 \vskip 0.4cm
{\it (5) Barlenses, manifestations of evolved bars:}

\noindent Barlenses are typically embedded in the inner parts 
of those bars that have ansae at the two ends of the bar
(ansae exist in 52$\pm$9$\%$ vs. 24$\pm$6$\%$ of barred galaxies with
and without barlenses, respectively), and which ansae in the simulation models
(Athanassoula $\&$ Misioritis 2002) are associated with evolved bars.
The frequency of barlenses does not depend on the galaxy brightness
(31$\pm$5$\%$ vs. 32$\pm$7$\%$ in the two galaxy luminosity bins,
respectively). Also, multiple lenses are rare in the galaxies with
barlenses (24$\pm$2$\%$ vs. 56$\pm$6$\%$ in barred galaxies with and without
barlenses, respectively).

 \vskip 0.4cm
{\it  (6) Nuclear bars:}

\noindent Contrary to the previous results we find that nuclear bars
appear more frequently in the weakly barred (AB) than in strongly
barred (B) galaxies (44$\pm$8$\%$ vs. 24$\pm$6$\%$ among the AB and B
galaxies, respectively), which is consistent with the theoretical
models by Maciejewski $\&$ Athanassoula (2008).  Contrary to the main
bars, nuclear bars (+ nuclear lenses) appear in a similar manner in
all Hubble types in NIRS0S (until the fraction drops at T=$-$3).
 \vskip 0.3cm

\noindent Nuclear sized single bars are detected in seven 'non-barred'
galaxies.  The small sizes of the bars are explained due to the large
bulges of these galaxies: they might have prevented the formation of
larger bars, due to the high $\Omega$ - $\kappa$/2 barrier for the bar pattern speed,
placing the co-rotation fairly close to the galaxy center.
 \vskip 0.4cm

{\it (7) Does the formation of the structure depend on the galaxy luminosity:}

\noindent The family classes A, AB and B cover the same galaxy
luminosity range, but the mean luminosity gradually decreases from A
-$>$ AB -$>$ B, so that the non-barred galaxies are on average 0.6
magnitudes more luminous than the strongly barred galaxies.  However, there
is no upper limit in galaxy luminosity for bar formation.  Also,
galaxy luminosity does not uniquely define the sizes of bars, or the
structures related to bars.
\vskip 0.4cm

\section*{Acknowledgments}

We acknowledge the significant observing time allocated to this project
during 2003-2009, based on observations made with several telescopes.
They include the New Technology Telescope (NTT), operated at the
Southern European Observatory (ESO), William Herschel Telescope (WHT),
the Italian Telescopio Nazionale Galileo (TNG), and the Nordic Optical
Telescope (NOT), operated on the island of La Palma. 
EL, HS, EA and AB acknowledge financial support to the DAGAL network from 
the People Programme (Marie Curie Actions)
of the European Union?s Seventh Framework Programme FP7/2007- 2013/ under 
REA grant agreement number PITN-GA-2011-289313.

\clearpage
\newpage

\begin{table}
\begin{center}
\caption{Fractions of structure components in different Hubble types
  in the {\it complete NIRS0S}(S0-sa). multi$_{l}$ contains 
  full lenses, and multi$_{l,rl}$ contains both lenses and ringlenses.
multi$_{bar}$ means galaxies with two or more bars identified visually,
whereas multi$_{bar}$(res) includes also weak bars detected by other means.   
N$_{tot}$ is the total galaxy number in each Hubble type bin, and the numbers 
for the different structure component are 
given in parenthesis. The Hubble types
earlier than T= $-$3 are not shown. \label{tbl-1}}
\begin{tabular}{llllll}
\hline
  \noalign{\smallskip}   
     \noalign{\smallskip}   
 Hubble stage: &   $-$3         & $-$2             & $-$1            & 0            & 1                \\
 \noalign{\smallskip}   
 (N$_{tot}$ in bin) &   (49)           &  (40)              &  (32)             & (29)           & (33)    \\
 \noalign{\smallskip}   
 \noalign{\smallskip}  
\hline
 \noalign{\smallskip}  
  \noalign{\smallskip}   
 {\it Bars:}              &                &                &               &               &                 \\
 \noalign{\smallskip}   
\noalign{\smallskip}   
A             & 63$\pm$7$\%$ (31)   & 50$\pm$8$\%$ (20)    & 16$\pm$6$\%$ (5)   &24$\pm$8$\%$ (7)    &18$\pm$7$\%$ (6)       \\
 \noalign{\smallskip}  
  \noalign{\smallskip}   
AB            &  20$\pm$6$\%$ (10)   &  20$\pm$6$\%$ (8)  &  41$\pm$9$\%$ (13)  & 38$\pm$9$\%$ (11)  & 51$\pm$9$\%$ (17)      \\
 \noalign{\smallskip} 
  \noalign{\smallskip}    

B             & 14$\pm$5$\%$ (7)    & 30$\pm$7$\%$ (12)    & 44$\pm$9$\%$ (14)   &38$\pm$9$\%$ (11)    &30$\pm$8$\%$ (10)       \\
 \noalign{\smallskip} 
  \noalign{\smallskip}     
B+AB          & 35$\pm$7$\%$ (17)    & 50$\pm$8$\%$ (20)    & 84$\pm$6$\%$ (27)   &76$\pm$8$\%$ (22)    &82$\pm$7$\%$ (27)    \\
 \noalign{\smallskip}  
  \noalign{\smallskip}   
multi$_{bar}$      & 6$\pm$3$\%$ (3) & 10$\pm$5$\%$ (4) & 22$\pm$7$\%$ (7) & 14$\pm$6$\%$ (4)  & 9$\pm$5$\%$ (3)          \\
 \noalign{\smallskip}
   \noalign{\smallskip}    
multi$_{bar}$(res) &  10$\pm$4$\%$ (5)  & 15$\pm$6$\%$ (6) &37$\pm$9$\%$ (12)  & 17$\pm$7$\%$ (5) & 21$\pm$7$\%$ (7)       \\
 \noalign{\smallskip}   
\noalign{\smallskip}   
 \noalign{\smallskip}
\hline
  \noalign{\smallskip}
  \noalign{\smallskip}  
{\it Lenses:}           &            &           &          &         &               \\    
 \noalign{\smallskip}  
 \noalign{\smallskip}   
L             & 26$\pm$6$\%$ (13)    & 18$\pm$6$\%$ (7)   &9$\pm$5$\%$ (3)  & 3$\pm$3$\%$ (1)    & 3$\pm$3$\%$ (1)         \\
 \noalign{\smallskip}  
 \noalign{\smallskip}    
l             & 28$\pm$6$\%$ (14)  & 25$\pm$7$\%$ (10)  & 9$\pm$5$\%$ (3)  & 14$\pm$6$\%$ (4)   & 12$\pm$6$\%$ (4)       \\
 \noalign{\smallskip}  
  \noalign{\smallskip}   
nl            & 8$\pm$4$\%$ (4)  & 20$\pm$6$\%$ (8)  & 22$\pm$7$\%$ (7)  & 21$\pm$8$\%$ (6)  & 18$\pm$7$\%$ (6)        \\
 \noalign{\smallskip}
  \noalign{\smallskip}     
bl            &  6$\pm$3$\%$ (3)  & 18$\pm$6$\%$ (7)  & 38$\pm$9$\%$ (12)  & 24$\pm$8$\%$ (7)  & 21$\pm$7$\%$ (7)       \\
 \noalign{\smallskip} 
 \noalign{\smallskip} 
multi$_{l}$   &  14$\pm$5$\%$ (7)  & 15$\pm$6$\%$ (6)  & 6$\pm$4$\%$ (2)  & 7$\pm$5$\%$ (2)  &6$\pm$4$\%$ (2)     \\  
 \noalign{\smallskip} 
 \noalign{\smallskip}    
L,RL          &  31$\pm$7$\%$ (15)   & 35$\pm$7$\%$ (14)  & 34$\pm$8$\%$ (11)  & 34$\pm$9$\%$ (10)  & 12$\pm$6$\%$ (4)    \\
 \noalign{\smallskip} 
  \noalign{\smallskip}    
  RL          &   4$\pm$2$\%$ (21)   & 18$\pm$6$\%$ (7)  & 25$\pm$7$\%$ (8)   & 31$\pm$9$\%$ (9)  & 9$\pm$5 (3)           \\
 \noalign{\smallskip} 
  \noalign{\smallskip}    

l,rl          &  28$\pm$6$\%$ (14)   & 40$\pm$8$\%$ (16)  &  41$\pm$9$\%$ (13)  & 28$\pm$8$\%$ (8)   & 21$\pm$7$\%$ (7)     \\
 \noalign{\smallskip} 
  \noalign{\smallskip}    

 rl          &     -      & 15$\pm$6$\%$ (6)  & 31$\pm$ 8$\%$ (10)   & 14$\pm$ 6$\%$ (4)  & 9$\pm$ 6$\%$ (3)      \\
 \noalign{\smallskip} 
  \noalign{\smallskip}    

nl,rnl        &  10$\pm$4$\%$ (5)   & 25$\pm$7$\%$ (10)  & 25$\pm$8$\%$ (8)  & 24$\pm$8$\%$ (7)  & 18$\pm$7$\%$ (6)     \\
 \noalign{\smallskip} 
  \noalign{\smallskip}   
multi$_{l,rl}$     &  16$\pm$5$\%$ (8)  & 30$\pm$7$\%$ (12)  &  28$\pm$8$\%$ (9)  &  24$\pm$8$\%$ (7)  &  9$\pm$5$\%$ (3)    \\
 \noalign{\smallskip}  
 \noalign{\smallskip}
 \hline
   \noalign{\smallskip}   
   \noalign{\smallskip} 
{\it Rings:}            &            &           &          &         &                \\
 \noalign{\smallskip} 
  \noalign{\smallskip}   
outer         &     -     &  5$\pm$3$\%$ (2)  & 25$\pm$8$\%$ (8)  &  21$\pm$8$\%$ (6)  & 42$\pm$9$\%$ (14)    \\
 \noalign{\smallskip}  
  \noalign{\smallskip}   
inner(r,rs)   &   2$\pm$2$\%$ (1)   & 8$\pm$4$\%$ (3)   & 28$\pm$8$\%$ (9)   & 21$\pm$8$\%$ (6)   & 15$\pm$6$\%$ (5)      \\
 \noalign{\smallskip}  
  \noalign{\smallskip}   
nuclear       &      -     & 3$\pm$3$\%$ (1)  & 25$\pm$8$\%$ (8)   & 7$\pm$5$\%$ (2)  & 12$\pm$6$\%$ (4)      \\
 \noalign{\smallskip}  
\hline

\end{tabular}
\end{center}
\end{table}


\clearpage
\newpage

\begin{table}
\begin{center}
\caption{Fractions of structure components in $B/T$-bins, for {\it complete NIRS0S}.  \label{tbl-2}}
\begin{tabular}{lllll}
\hline

 \noalign{\smallskip}
  \noalign{\smallskip}        
 $B/T$-bin: &  0 - 0.2   & 0.2 - 0.3 &  0.3 - 0.4 & $>$ 0.4           \\
 \noalign{\smallskip} 
 (N$_{tot}$ in bin) &  (55)     & (56)        &  (41)        &   (53)       \\
 \noalign{\smallskip}   
\noalign{\smallskip}   
\hline
 \noalign{\smallskip}  
  \noalign{\smallskip}  
{\it Bars:}         &            &           &     &                         \\
  \noalign{\smallskip}   
\noalign{\smallskip}   
A            & 38$\pm$7$\%$ (21) & 30$\pm$6$\%$ (17) & 37$\pm$8$\%$ (15) &  66$\pm$7$\%$ (35)   \\      
 \noalign{\smallskip}  
  \noalign{\smallskip}   
AB           & 22$\pm$6$\%$ (12) & 39$\pm$7$\%$ (22) & 44$\pm$8$\%$ (18) &   15$\pm$5$\%$ (8)   \\    
 \noalign{\smallskip} 
  \noalign{\smallskip}  
B            & 40$\pm$7$\%$ (22) & 29$\pm$6$\%$ (16) & 17$\pm$6$\%$ (7)&   17$\pm$5$\%$ (9)    \\      
 \noalign{\smallskip} 
  \noalign{\smallskip}        
B+AB         &64$\pm$6$\%$ (35) & 68$\pm$6$\%$ (38) & 61$\pm$8$\%$ (25) &  32$\pm$6$\%$ (17)     \\      
 \noalign{\smallskip}  
  \noalign{\smallskip}   
 \noalign{\smallskip}
  \noalign{\smallskip}
    \noalign{\smallskip}
{\it Lenses:}      & & & &                                        \\                  
 \noalign{\smallskip}  
 \noalign{\smallskip}   
L            & 9$\pm$ (5) & 11$\pm$4$\%$ (6) & 22$\pm$6$\%$ (9) & 9$\pm$4$\%$ (5)   \\      
 \noalign{\smallskip}  
 \noalign{\smallskip}    
l            & 7$\pm$4 (4) & 14$\pm$5$\%$ (8) &  24$\pm$7$\%$ (10)& 26$\pm$6$\%$ (14) \\      
 \noalign{\smallskip}  
  \noalign{\smallskip}   
nl           & 9$\pm$4 (5) & 21$\pm$5$\%$ (12) & 20$\pm$6$\%$ (8) & 11$\pm$4$\%$ (6)  \\      
 \noalign{\smallskip}
  \noalign{\smallskip}     
bl           & 24$\pm$6 (13) &29$\pm$6$\%$ (16) & 15$\pm$5$\%$ (6) & 2$\pm$2$\%$ (1)  \\      
 \noalign{\smallskip} 
   \noalign{\smallskip}   

 \noalign{\smallskip} 
 \noalign{\smallskip}    
L,RL         & 24$\pm$6 (13) & 34$\pm$6$\%$ (19) & 34$\pm$7$\%$ (14) &  15$\pm$5$\%$ (8)  \\      
 \noalign{\smallskip} 
  \noalign{\smallskip}    
RL          & 15$\pm$5 (8)   & 23$\pm$6$\%$ (13) & 12$\pm$5$\%$ (5) & 6$\pm$3$\%$ (3)  \\
 \noalign{\smallskip} 
  \noalign{\smallskip}    
l,rl         & 18$\pm$ (10) &30$\pm$6$\%$ (17) & 41$\pm$7$\%$ (17)& 32$\pm$6$\%$ (17)  \\      
 \noalign{\smallskip} 
  \noalign{\smallskip}    
rl           & 11$\pm$4 (6) & 16$\pm$5$\%$ (9) 1& 17$\pm$6$\%$ (7) & 6$\pm$3$\%$ (3)  \\
 \noalign{\smallskip} 
  \noalign{\smallskip} 
nl,rnl       &11$\pm$4 (6) &23$\pm$6$\%$ (13) & 20$\pm$6$\%$ (8)& 17 $\pm$5$\%$ (9)   \\      
 \noalign{\smallskip} 
   \noalign{\smallskip}
    \noalign{\smallskip} 
{\it Rings:}       & & & &             \\      
 \noalign{\smallskip} 
  \noalign{\smallskip}   
outer       &16$\pm$5 (9) & 23$\pm$6$\%$(13)& 12$\pm$5$\%$ (5)& 6$\pm$3$\%$ (3)  \\       
 \noalign{\smallskip}  
  \noalign{\smallskip}   
in(r,rs)  &16$\pm$5 (9) &14$\pm$5$\%$ (8) &7 $\pm$4$\%$ (3) & 7$\pm$4$\%$ (4)   \\      
 \noalign{\smallskip}  
  \noalign{\smallskip}   
nuclear      &9$\pm$4 (5) & 9$\pm$4$\%$ (5)& 12$\pm$5$\%$ (5)&  -        \\      
 \noalign{\smallskip}  
\hline

\end{tabular}
\end{center}
\end{table}

\clearpage
\newpage
\begin{table}
\begin{center}
\caption{Fractions of structure components in bins of absolute galaxy magnitudes in $K$-band, for {\it complete NIRS0S}. \label{tbl-3}}
\begin{tabular}{llll}
\hline
 \noalign{\smallskip}
  \noalign{\smallskip}        
 Magnitude bin:  & $-$20 to $-$22  & $-$22 to $-$24 & $-$24 to $-$26           \\
\noalign{\smallskip}
 (N$_{tot}$ in bin) &  (7)  & (96)             &  (102)                              \\
 \noalign{\smallskip}
  \noalign{\smallskip}  
\hline
 \noalign{\smallskip}
 \noalign{\smallskip}   
{\it  Bars:}   &                  &           &                        \\
 \noalign{\smallskip}  
\noalign{\smallskip}   
A             & (2) & 26$\pm$5$\%$ (25)   &  59$\pm$5$\%$ (60)    \\
 \noalign{\smallskip}  
  \noalign{\smallskip}   
AB            & (1) &  36$\pm$5$\%$ (35)     &  24$\pm$4$\%$ (24)    \\
 \noalign{\smallskip} 
  \noalign{\smallskip} 
B             & (4)  &  35$\pm$5$\%$ (34)  &  16$\pm$4$\%$ (16)  \\
 \noalign{\smallskip} 
  \noalign{\smallskip}     
B+AB         & (5) & 73$\pm$5$\%$ (70)  & 39$\pm$5$\%$ (40)    \\
 \noalign{\smallskip}  
  \noalign{\smallskip}   
nb(res)      & (-) &  26$\pm$4$\%$ (25) &  18$\pm$4$\%$ (18)  \\

  \noalign{\smallskip}   
\noalign{\smallskip}   
  \noalign{\smallskip}
  \noalign{\smallskip}  
{\it Lenses:}   &                 &             &                 \\    
 \noalign{\smallskip}  
 \noalign{\smallskip}   
L           & (2)  & 14$\pm$4$\%$ (13) & 10$\pm$3$\%$ (10)   \\
 \noalign{\smallskip}  
 \noalign{\smallskip}    
l           & (-) &  16$\pm$4$\%$ (15) & 21$\pm$4$\%$ (21)   \\
 \noalign{\smallskip}  
  \noalign{\smallskip}   
nl         & (1)  &  14$\pm$3$\%$ (13) &  17$\pm$4$\%$ (17)   \\
 \noalign{\smallskip}
  \noalign{\smallskip}     
bl         & (2)  & 25$\pm$4$\%$ (24) & 10$\pm$3$\%$ (10)      \\
 \noalign{\smallskip} 
   \noalign{\smallskip}   
multi$_{l}$ &(1) & 10$\pm$3$\%$ (9)  & 8$\pm$3$\%$ (8)  \\
 \noalign{\smallskip} 
 \noalign{\smallskip}    
L,RL        & (4) &  32$\pm$5$\%$ (31) &  19$\pm$4$\%$ (19)  \\
 \noalign{\smallskip} 
  \noalign{\smallskip}    
l,rl        & (-) &  31$\pm$5$\%$ (30)   & 31$\pm$4$\%$ (31)   \\
 \noalign{\smallskip} 
  \noalign{\smallskip}    
nl,rnl      & (1) & 18$\pm$4$\%$ (17)  & 18$\pm$4$\%$ (18)    \\
 \noalign{\smallskip} 
  \noalign{\smallskip}   
multi$_{l,rl}$ & (1)  & 22$\pm$4$\%$ (21) & 17$\pm$4$\%$ (17)   \\
 \noalign{\smallskip} 
  \noalign{\smallskip}
   \noalign{\smallskip}  
{\it Rings:}   &                &             &        \\
 \noalign{\smallskip} 
  \noalign{\smallskip}   
outer       & (1) & 18$\pm$4$\%$ (17)  &  12$\pm$3$\%$ (12)   \\
 \noalign{\smallskip}  
  \noalign{\smallskip}   
inner(r,rs) & (2) & 20$\pm$4$\%$ (19)& 3$\pm$2$\%$ (3)    \\
 \noalign{\smallskip}  
  \noalign{\smallskip}   
nuclear     &(1) & 10$\pm$3$\%$ (10) & 4$\pm$2$\%$ (4)    \\
 \noalign{\smallskip}  
\hline

\end{tabular}
\end{center}
\end{table}

\clearpage
\newpage


\begin{table}
\begin{center}
\caption{Fractions of structure components in different Hubble types in {\it barred galaxies} in the complete NIRS0S. \label{tbl-4}}
\begin{tabular}{llllll}
\hline
 \noalign{\smallskip} 
 Barred      &                    &                &               &               &                 \\
 \noalign{\smallskip}         
 Hubble stage: &   $-$3         & $-$2             & $-$1            & 0            & 1              \\
 \noalign{\smallskip} 
 (N$_{tot}$ in bin)     & (17)         &    (20)            &  (27)             & (22)           & (27)       \\
 \noalign{\smallskip}   
 \noalign{\smallskip}  
\hline
 \noalign{\smallskip}
 \noalign{\smallskip}  
{\it Bars:}       &                    &                &               &               &                  \\
 \noalign{\smallskip}                                                                                  
 \noalign{\smallskip}
nb              & 29$\pm$11$\%$ (5)  & 35$\pm$11$\%$ (7)  & 26$\pm$8$\%$ (7) & 23$\pm$9$\%$ (5) & 11$\pm$6$\%$ (3)   \\
 \noalign{\smallskip}   
 \noalign{\smallskip}   
nb+nb$_{res}$     & 41$\pm$12$\%$ (7)  & 45$\pm$11$\%$ (9)  &48$\pm$10$\%$ (13)  & 27$\pm$9$\%$ (6) & 29$\pm$9$\%$ (8)    \\
 \noalign{\smallskip}  
 \noalign{\smallskip}    
B$_a$           & 12$\pm$8$\%$ (2) & 30$\pm$10$\%$ (6) & 19$\pm$7$\%$ (5)  & -  & 14$\pm$7$\%$ (4)  \\
 \noalign{\smallskip} 
   \noalign{\smallskip}   
AB$_a$           & 18$\pm$9$\%$ (3)  & 20$\pm$9$\%$ (4) & 19$\pm$7$\%$ (5) & 18$\pm$8$\%$ (4) & 7$\pm$5$\%$ (2)   \\
 \noalign{\smallskip}
  \noalign{\smallskip}     
B$_{ax}$         &  11$\pm$7$\%$ (2) & 40$\pm$10$\%$ (8) & 22$\pm$8$\%$ (6) & 9$\pm$6$\%$ (2)   & 19$\pm$7$\%$ (5)      \\
 \noalign{\smallskip} 
   \noalign{\smallskip}   
AB$_{ax}$        &  17$\pm$8$\%$ (3) & 20$\pm$9$\%$ (4) &  26$\pm$8$\%$ (7) & 18$\pm$8$\%$ (4)  &  15$\pm$7$\%$ (4)      \\
 \noalign{\smallskip}
    \noalign{\smallskip}   
BAB$_{ax}$       &  29$\pm$11$\%$ (5) & 60$\pm$11$\%$ (12) & 48$\pm$10$\%$ (13) & 27$\pm$9$\%$ (6) &  33$\pm$9$\%$ (9)     \\
 \noalign{\smallskip} 
   \noalign{\smallskip}   
multi$_{bar}$       & 18$\pm$9$\%$ (3) &  20$\pm$9$\%$ (4) &  26$\pm$8$\%$ (7) &  18$\pm$8$\%$ (4) & 11$\pm$6$\%$ (3)      \\
 \noalign{\smallskip} 
   \noalign{\smallskip}   
multi$_{bar}$(res)   & 29$\pm$11$\%$ (5) &  30$\pm$10$\%$ (6) & 44$\pm$10$\%$ (12) &  23$\pm$9$\%$ (5) & 26$\pm$8$\%$ (7)    \\
 \noalign{\smallskip}   
 \noalign{\smallskip}
 \noalign{\smallskip}  
{\it Lenses:}            &                &                &               &               &                 \\
 \noalign{\smallskip}  
 \noalign{\smallskip}
L               &  59$\pm$12$\%$ (10) & 25$\pm$9$\%$ (5) & 11$\pm$6$\%$ (3) &  4$\pm$4$\%$ (1) & 4$\pm$4$\%$ (1)      \\
 \noalign{\smallskip}
    \noalign{\smallskip}   
l                &  41$\pm$12$\%$ (7)  & 20$\pm$9$\%$ (4) & 11$\pm$6$\%$ (3)  &  5$\pm$4$\%$ (1) &  11$\pm$6$\%$ (3)       \\
 \noalign{\smallskip}
    \noalign{\smallskip}   
nl              &  6$\pm$6$\%$ (1)  & 20$\pm$9$\%$ (4) & 22$\pm$8$\%$ (6) &  18$\pm$8$\%$ (4) & 18$\pm$7$\%$ (5)     \\
 \noalign{\smallskip}
    \noalign{\smallskip}   
bl               & 18$\pm$9$\%$ (3) & 45$\pm$11$\%$ (9) &  41$\pm$9$\%$ (11) &  32$\pm$10$\%$ (7)  & 30$\pm$9$\%$ (8)      \\
 \noalign{\smallskip}
    \noalign{\smallskip}   
multi$_{l}$     & 18$\pm$9$\%$ (3) & 15$\pm$8$\%$ (3) &  7$\pm$5$\%$ (2) &  -    & 7$\pm$5$\%$ (2)     \\
 \noalign{\smallskip}  
 \noalign{\smallskip}   
L,RL            &  59$\pm$12$\%$ (10)  &  40$\pm$11$\%$ (8)  & 37$\pm$9$\%$ (10) & 36$\pm$10$\%$ (8) & 11$\pm$6$\%$ (3)     \\
 \noalign{\smallskip} 
   \noalign{\smallskip}   
RL                & -     & 15$\pm$8$\%$ (3)    & 26$\pm$8$\%$ (7) & 32 $\pm$10$\%$ (7) & 7$\pm$5$\%$ (2)     \\
 \noalign{\smallskip} 
   \noalign{\smallskip}   
l,rl             &  41$\pm$12$\%$ (7) & 25$\pm$10$\%$ (5) &  41$\pm$9$\%$ (11) & 23$\pm$9$\%$ (5) & 22$\pm$8$\%$ (6)     \\
 \noalign{\smallskip}  
  \noalign{\smallskip}   
rl                 & -   &  5$\pm$5$\%$ (1) & 30$\pm$9$\%$ (8)  & 18$\pm$8$\%$ (4) & 11$\pm$6$\%$ (3)     \\
 \noalign{\smallskip}  
  \noalign{\smallskip}   

nl,rnl          &  12$\pm$8$\%$ (2)  & 30$\pm$10$\%$ (6) &26$\pm$8$\%$ (7)  &  23$\pm$9$\%$ (5) &  19$\pm$7$\%$ (5)     \\
 \noalign{\smallskip} 
   \noalign{\smallskip}   
multi$_{l,rl}$     & 24$\pm$10$\%$ (4) &  30$\pm$10$\%$ (6)  & 30$\pm$9$\%$ (8)  & 18$\pm$8$\%$ (4)&  11$\pm$6$\%$ (3)    \\
 \noalign{\smallskip}  
 \noalign{\smallskip} 
  \noalign{\smallskip} 
{\it Rings:}             &              &                  &               &              &                       \\
 \noalign{\smallskip} 
  \noalign{\smallskip}   
outer            &    -    &  10$\pm$7$\%$ (2)  &  30$\pm$9$\%$ (8)  & 23$\pm$9$\%$ (5) & 44$\pm$10$\%$ (12)       \\
 \noalign{\smallskip}  
  \noalign{\smallskip}   
inner(r,rs)      &        -      &   10$\pm$7$\%$ (2)  & 33$\pm$9$\%$ (9)  & 18$\pm$8$\%$ (4)  & 11$\pm$6$\%$ (3)      \\
 \noalign{\smallskip} 
   \noalign{\smallskip}   
nuclear          &         -       &  5$\pm$5$\%$ (1)    &  30$\pm$9$\%$ (8)   & 9$\pm$6$\%$ (2)  &  11$\pm$6$\%$ (3)       \\
 \noalign{\smallskip}  
\hline

\end{tabular}
\end{center}
\end{table}

\clearpage
\newpage


\begin{table}
\begin{center}
\caption{Fractions of structure components in $B/T$-bins in {\it barred galaxies} in the complete NIRS0S. \label{tbl-5}}
\begin{tabular}{lllll}
\hline
 \noalign{\smallskip}  
 Barred & & & & \\      
 \noalign{\smallskip}         
 $B/T$-bin: &    0 - 0.2 & 0.2 - 0.3 & 0.3 - 0.4 & $>$ 0.4       \\
 \noalign{\smallskip}  
 (N$_{tot}$ in bin)     & (35)  & (37)        & (25)        &  (18) \\
 \noalign{\smallskip}   
 \noalign{\smallskip}  
\hline
 \noalign{\smallskip}
  \noalign{\smallskip} 
{\it Bars:}        & & & &            \\      
 \noalign{\smallskip}                                                                                  
 \noalign{\smallskip}             
nb           &11$\pm$5$\%$ (4) & 22$\pm$7$\%$ (8) & 36$\pm$10$\%$ (9) & 17$\pm$9$\%$ (3)  \\        
 \noalign{\smallskip}   
 \noalign{\smallskip}   
nb+nb$_{r}$   &20$\pm$7$\%$ (7) & 38$\pm$8$\%$ (14)& 44$\pm$10$\%$ (11) & 17$\pm$9$\%$ (3)   \\       
 \noalign{\smallskip}  
 \noalign{\smallskip}    
B$_a$        & 20$\pm$7$\%$ (7) & 16$\pm$6$\%$ (6) &4$\pm$4$\%$ (1) & 17$\pm$9$\%$ (3)     \\                 
 \noalign{\smallskip} 
   \noalign{\smallskip}   
AB$_a$       & 9$\pm$5$\%$  (3) & 22$\pm$7$\%$  (8)& 24$\pm$9$\%$  (6)& 6$\pm$ 5$\%$  (1)      \\                
 \noalign{\smallskip}
  \noalign{\smallskip}     
B$_{ax}$    & 23$\pm$7$\%$  (8) &22$\pm$7$\%$  (8) &8$\pm$5$\%$  (2) & 28$\pm$11$\%$  (5)     \\               
 \noalign{\smallskip} 
   \noalign{\smallskip}   
AB$_{ax}$  &11$\pm$5$\%$  (4) & 24$\pm$7$\%$  (9) &24$\pm$8$\%$  (6) & 17$\pm$9$\%$  (3)       \\               
 \noalign{\smallskip}
    \noalign{\smallskip}   
BAB$_{ax}$  &34$\pm$8$\%$  (12) & 46$\pm$8$\%$  (17)& 32$\pm$9$\%$  (8)& 44$\pm$12$\%$  (8)     \\              
 \noalign{\smallskip} 
   \noalign{\smallskip}   
multi$_{bar}$ &20$\pm$7$\%$  (7) & 38$\pm$8$\%$  (14) & 44$\pm$10$\%$  (11) & 17$\pm$9$\%$  (3)    \\
 \noalign{\smallskip} 
   \noalign{\smallskip}   
  \noalign{\smallskip} 
{\it Lenses:}     & & & &         \\
 \noalign{\smallskip}  
 \noalign{\smallskip}
L          &14$\pm$6$\%$  (5) &13$\pm$6$\%$  (5) &32$\pm$9$\%$  (8) &  11$\pm$7$\%$  (2)  \\             
 \noalign{\smallskip}
    \noalign{\smallskip}   
l          &3$\pm$3$\%$  (1) &8$\pm$5$\%$  (3) &28$\pm$9$\%$  (7) & 33$\pm$11$\%$  (6)  \\           
 \noalign{\smallskip}
    \noalign{\smallskip}   
nl         &6$\pm$4$\%$  (2) &24$\pm$7$\%$  (9) & 28$\pm$9$\%$  (7)& 17$\pm$9$\%$  (3)     \\           
 \noalign{\smallskip}
    \noalign{\smallskip}   
bl         &37$\pm$8$\%$  (13) &43$\pm$8$\%$  (16) &24$\pm$8$\%$  (6) & 6$\pm$5$\%$  (1)    \\          
 \noalign{\smallskip}
    \noalign{\smallskip}   
multi$_{l}$  & 8$\pm$5$\%$  (3) & 8$\pm$5$\%$  (3)  & 28$\pm$9$\%$  (7) & 33$\pm$11$\%$  (6)   \\     
 \noalign{\smallskip}  
 \noalign{\smallskip}   
L,RL       &31$\pm$8$\%$  (11) &40$\pm$8$\%$  (15) &48$\pm$10$\%$  (12) & 22$\pm$10$\%$  (4)   \\         
 \noalign{\smallskip} 
   \noalign{\smallskip}   
RL         & 17$\pm$6$\%$  (6) & 27$\pm$7$\%$  (10) & 16$\pm$7$\%$  (4) & 11$\pm$7$\%$  (2)  \\
 \noalign{\smallskip} 
   \noalign{\smallskip}   
l,rl       &17$\pm$6$\%$  (6) &27$\pm$7$\%$  (10) &48$\pm$10$\%$  (12)  &  44$\pm$12$\%$  (8)  \\         
 \noalign{\smallskip}  
  \noalign{\smallskip}   
rl        & 14$\pm$6$\%$  (5) & 19$\pm$6$\%$  (7) & 20$\pm$8$\%$  (5) & 11$\pm$7$\%$  (2)     \\
 \noalign{\smallskip}  
  \noalign{\smallskip}   

nl,rnl     &9$\pm$5$\%$  (3) & 27$\pm$7$\%$  (10)&28$\pm$9$\%$  (7) &  33$\pm$11$\%$  (6)   \\              
 \noalign{\smallskip} 
   \noalign{\smallskip}   
multi$_{rl}$   &  9$\pm$5$\%$  (3)  &  24$\pm$7$\%$  (9)   & 20$\pm$8$\%$  (5)   & 33$\pm$11$\%$  (6)   \\      
 \noalign{\smallskip}  
 \noalign{\smallskip}
multi$_{l,rl}$   & 17$\pm$6$\%$  (6) & 32$\pm$8$\%$  (12)    & 48$\pm$19$\%$  (12) &  50$\pm$12$\%$  (9)     \\
 \noalign{\smallskip}  
{\it Rings:}     & & & &          \\
 \noalign{\smallskip} 
  \noalign{\smallskip}   
outer      &23$\pm$7$\%$  (8) & 35$\pm$8$\%$  (13)& 16$\pm$7$\%$  (4)& 11$\pm$7$\%$  (2)    \\          
 \noalign{\smallskip}  
  \noalign{\smallskip}   
in(r,rs)  &17$\pm$6$\%$  (6) & 19$\pm$6$\%$  (7)& 12$\pm$6$\%$ (3)& 11$\pm$7$\%$  (2)   \\            
 \noalign{\smallskip} 
   \noalign{\smallskip}   
nuclear      &14$\pm$6$\%$  (5) &11$\pm$5$\%$  (4) &20$\pm$8$\%$  (5) & -     \\              
 \noalign{\smallskip}  
\hline

\end{tabular}
\end{center}
\end{table}

\clearpage
\newpage

\begin{table}
\begin{center}
\caption{Fractions of structure components in bins of absolute galaxy brightness in $K$-band, for {\it barred galaxies} in the complete NIRS0S. \label{tbl-6}}
\begin{tabular}{llll}
\hline
 \noalign{\smallskip} 
 Barred & & & \\      
 \noalign{\smallskip}         
 Magnitude bin:& $-$20 to $-$22  & $-$22 to $-$24 & $-$24 to $-$26      \\
 \noalign{\smallskip} 
 (N$_{tot}$ in bin)&  (5)   & (70)            &  (40)  \\
 \noalign{\smallskip}   
 \noalign{\smallskip}  
\hline
 \noalign{\smallskip}
  \noalign{\smallskip} 
{\it Bars:}  &                      &          &          \\      
 \noalign{\smallskip}                                                                                  
 \noalign{\smallskip}             
nb           & (-) & 23$\pm$5$\%$ (16)  & 15$\pm$6$\%$ (6)  \\     
 \noalign{\smallskip}   
 \noalign{\smallskip}   
nb+nb$_{res}$ & (-) & 31$\pm$5$\%$ (22)   & 33$\pm$7$\%$ (13) \\     
 \noalign{\smallskip}  
 \noalign{\smallskip}    
B$_a$        & (1) & 14$\pm$4$\%$ (10)   & 15$\pm$6$\%$ (6) \\     
 \noalign{\smallskip} 
   \noalign{\smallskip}   
AB$_a$       & (0) & 19$\pm$5$\%$ (13)   &  13$\pm$5$\%$ (5) \\     
 \noalign{\smallskip}
  \noalign{\smallskip}     
B$_{ax}$      &(1)  & 20$\pm$5$\%$ (14)  & 20$\pm$6$\%$ (8)  \\     
 \noalign{\smallskip} 
   \noalign{\smallskip}   
AB$_{ax}$     & (-) & 23$\pm$5$\%$ (16)  & 15$\pm$6$\%$ (6)  \\     
 \noalign{\smallskip}
    \noalign{\smallskip}   
BAB$_{ax}$    & (1) & 43$\pm$6$\%$ (30)  & 38$\pm$8$\%$ (15) \\     
 \noalign{\smallskip} 
   \noalign{\smallskip}   
multi$_{bar}$  & (-)  &  31$\pm$6$\%$ (22)   &    33$\pm$7$\%$ (13)       \\
 \noalign{\smallskip}  
 \noalign{\smallskip}
 \noalign{\smallskip}  
{\it Lenses:}   &                &          &            \\     
 \noalign{\smallskip}  
 \noalign{\smallskip}
L              & (1)  & 19$\pm$5$\%$ (13)  & 15$\pm$6$\%$ (6) \\     
 \noalign{\smallskip}
    \noalign{\smallskip}   
l              & (0)  & 17$\pm$5$\%$ (12)  & 13$\pm$5$\%$ (5)   \\     
 \noalign{\smallskip}
    \noalign{\smallskip}   
nl             & (0)  & 16$\pm$4$\%$ (11)  & 25$\pm$7$\%$ (10)  \\     
 \noalign{\smallskip}
    \noalign{\smallskip}   
bl            & (2)   & 34$\pm$6$\%$ (24)  & 25$\pm$6$\%$ (10)   \\     
 \noalign{\smallskip}
    \noalign{\smallskip}   
multi$_{l}$  &(1) &  14$\pm$5$\%$ (10)    &  20$\pm$6$\%$ (8)   \\     
 \noalign{\smallskip}  
L,RL          & (3)  & 40$\pm$6$\%$ (28) &  28$\pm$7$\%$ (11)  \\      
 \noalign{\smallskip} 
   \noalign{\smallskip}   

RL            & (1)  & 21$\pm$ 5$\%$ (15)  & 12$\pm$ 5$\%$ (5)   \\
 \noalign{\smallskip} 
   \noalign{\smallskip}   

l,rl          & (0)  & 36$\pm$6$\%$ (25) & 28$\pm$7$\%$ (11)  \\     
 \noalign{\smallskip}  
  \noalign{\smallskip}   
rl             & (0) & 18$\pm$5$\%$ (13)& 15$\pm$6$\%$ (6)  \\
 \noalign{\smallskip}  
  \noalign{\smallskip}   

nl,rnl         & (0)   & 21$\pm$ 5$\%$ (15) &  28$\pm$7$\%$ (11)   \\     
 \noalign{\smallskip} 
   \noalign{\smallskip}   
multi$_{l,rl}$  & (1) &  10$\pm$3$\%$ (7)     &   30$\pm$7$\%$ (12)     \\      
 \noalign{\smallskip} 
 \noalign{\smallskip}
 \noalign{\smallskip}  
{\it Rings:}   &               &          &       \\     
 \noalign{\smallskip} 
  \noalign{\smallskip}   
outer            & (1)  &  23$\pm$5$\%$ (16) &  25$\pm$10$\%$ (10) \\      
 \noalign{\smallskip}  
  \noalign{\smallskip}   
inner(r,rs)       & (2)  & 20$\pm$5$\%$ (14) &  5$\pm$3$\%$ (2)   \\     
 \noalign{\smallskip} 
   \noalign{\smallskip}   
nuclear        & (1)    &  13$\pm$4$\%$ (9) &  10$\pm$5$\%$ (4)  \\     
 \noalign{\smallskip}  
\hline

\end{tabular}
\end{center}
\end{table}

\clearpage
\newpage

\begin{table}
\begin{center}
\caption{Fractions of structure components among the barred (B+AB) S0+S0/a galaxies: compared are galaxies with (left column), 
and without barlenses (right column).
The x-shaped bars are included into barlenses, and the eight nuclear bars without any main bars do not form part of the statistics. \label{tbl-8}}
\begin{tabular}{lll}
\hline
\noalign{\smallskip}
Barred S0-S0/a: & & \\
\noalign{\smallskip}  
        & have bl      & no bl           \\
 \noalign{\smallskip}
 (N$_{tot}$ in bin)  & (28)   &   (57) \\
  \noalign{\smallskip}       
\hline  & &  \\
  \noalign{\smallskip}        
ansae   & 52$\pm$9$\%$ (15)     &  24$\pm$6$\%$(14)    \\
nb      & 45$\pm$9$\%$ (13)     &  38$\pm$6$\%$(22)    \\
nl      & 21$\pm$8 $\%$(6)      &  16$\pm$5$\%$(9)    \\
nl+nrl  & 24$\pm$8$\%$ (7)      &  37$\pm$6$\%$(21)    \\   
multi$_{l,rl}$ & 24$\pm$8$\%$ (7) & 56$\pm$7$\%$(32)  \\   
 \noalign{\smallskip}      
\hline
\end{tabular}
\end{center}
\end{table}



\begin{table}
\begin{center}
\caption{Fractions of inner, outer, and nuclear rings among the barred and non-barred galaxies. 
The barred galaxies include both strong (B) and weak (AB) bars (uses {\it complete NIRS0S}, e.g. S0-Sa types). \label{tbl-7}}
\begin{tabular}{lll}
\hline
  \noalign{\smallskip}        
Fraction of & in barred   & in non-barred    \\
 \noalign{\smallskip}      
 (N$_{tot}$ in bin)     & (115)  & (79)  \\
  \noalign{\smallskip}        
\hline
  \noalign{\smallskip}        
inner ring & 17$\pm$3$\%$  (20)  & 7$\pm$3$\%$  (6)  \\
outer ring & 14$\pm$3$\%$  (16)  & 4$\pm$2$\%$  (3)  \\
nuclear ring & 12$\pm$3$\%$  (14) & 1$\pm$1$\%$  (1)  \\
  \noalign{\smallskip}      
\hline
\end{tabular}
\end{center}
\end{table}


\begin{table}
\begin{center}
 \caption{Fractions of structure components in A, AB and B family classes for the complete NIRS0S.
For the S0-S0/a galaxies calculated are also the fractions of nuclear bars among the strongly (B) 
and weakly (AB) barred galaxies, as well as the fractions of barlenses in the same family classes.  
In parentheses the total numbers of galaxies in the different family classes are given.   \label{tbl-9}}
\begin{tabular}{llll}
\hline
\noalign{\smallskip}
\noalign{\smallskip}  
family            & A     &  AB    &   B            \\
  \noalign{\smallskip} 
  \noalign{\smallskip}       
\hline
  \noalign{\smallskip}        
complete NIRS0S: & & & \\
  \noalign{\smallskip} 
(N$_{tot}$ in bin) & (82)  & (57)   & (52)       \\
  \noalign{\smallskip}   
L     &  5 $\pm$2$\%$ (4)  &  16$\pm$5$\%$ (9)    & 19$\pm$4$\%$ (12)  \\
l     &  21$\pm$4$\%$ (17) &  21$\pm$5$\%$ (12)   & 16$\pm$4$\%$ (6)  \\
nl    &  10$\pm$3$\%$  (8)  &  23$\pm$6$\%$ (13)   & 20$\pm$4$\%$ (9)   \\
RL    &  6$\pm$3$\%$  (5)  &  23$\pm$6$\%$ (13)   & 21$\pm$4$\%$ (10) \\
rl    &  9$\pm$3$\%$  (2)  &  23$\pm$6$\%$ (13)   & 17$\pm$4$\%$ (6)  \\
nrl   &  -            (0)  &  4$\pm$2$\%$  (2)    & 5$\pm$2$\%$  (3)   \\
R     &  4$\pm$2$\%$  (3)  &  30$\pm$6$\%$ (17)   & 25$\pm$4$\%$ (10)  \\
r     &  6$\pm$3$\%$  (5)  &  12$\pm$4$\%$ (7)    & 17$\pm$4$\%$ (12) \\
nr    &  -            (0)  &  16$\pm$5$\%$ (9)    & 14$\pm$3$\%$ (6)   \\
 \noalign{\smallskip}  
 \noalign{\smallskip}
S0+S0/a: & & & \\
(N$_{tot}$ in bin) & (54) & (41)   & (42)    \\
     \noalign{\smallskip}  
bl      &    -      & 38$\pm$9$\%$ (11)    & 62$\pm$9$\%$ (18)    \\
no bl   &    -      & 54$\pm$6$\%$ (31)    & 45$\pm$6$\%$ (26)    \\
nb      &    13$\pm$5$\%$ (7)    & 44$\pm$8$\%$ (18)    & 24$\pm$6$\%$ (19)   \\
 \noalign{\smallskip}      
\hline 
\end{tabular}
\end{center}
\end{table}

\clearpage
\newpage


\begin{table}
\begin{center}
\caption{Nuclear bars in the non-barred galaxies. The numbers are given
in bins of the Hubble type T, $B/T$, and in the three magnitude bins. The bins 
are: bin1 = $-$22$\leq$M$_k$$<$$-$20, bin2 = $-$24$\leq$M$_k$$<$$-$22,
bin3 = $-$26$\leq$M$_k$$<$$-$24. \label{tbl-10}}
\begin{tabular}{ll}
  \noalign{\smallskip}       
\hline  
  \noalign{\smallskip}        
T-range: & $-$3 (2), $-$2 (2), $-$1 (1), 0 (1), 1 (1)      \\
Mag-range: & bin1 (0), bin2 (2), bin3 (5)      \\
$B/T$-range & 0-0.2 (0), 0.2-0.3 (1), 0.3-0.4 (2), $>$0.4 (4)      \\
 \noalign{\smallskip}      
\hline
\end{tabular}
\end{center}
\end{table}


\begin{table}
\begin{center}
\caption{{\it Galaxies with multiple lenses and multiple bars}. The mean ratios of the major axis radii (length) of inner and outer lenses are given.
In parenthesis are indicated whether only fully developed lenses (l,L) form part of the multiple lens,
or are ringlenses (rl,RL) also included. The last two
  values are obtained as peak-values in the the histograms of the size distributions of
  the structures, normalized to the bar size. The uncertainties are the standard deviations of the mean. \label{tbl-11}}
\begin{tabular}{lll}
 \noalign{\smallskip}        
\hline  &  & \\
 \noalign{\smallskip}  
{\it Multiple lenses:} & & \\
 \noalign{\smallskip}  
length(L,RL)/length(l,rl)   & 1.9$\pm$0.2 (N=8)  & barred \\
length(L,RL)/length(l,rl)   & 2.5$\pm$0.2 (N=8) & non-barred \\
  \noalign{\smallskip}
      \noalign{\smallskip}     
length(L)/length(l)         & 2.85$\pm$0.19 (N=4) & all \\
length(L,RL)/length(l,rl)   & 1.96$\pm$0.14 (N=11) & all \\
  \noalign{\smallskip}
 length(L,RL)/length(bl)    & 3.8           & barred      \\
 length(outer-ring)=length(inner-ring) & 2.0  & barred       \\
  \noalign{\smallskip}      
  \noalign{\smallskip}     
{\it Multiple bars:} & & \\
  \noalign{\smallskip} 
length(bar)/length(nb)   & 8.0$\pm$1.0 (N=23) & weak (AB) bars \\
length(bar)/length(nb)   & 6.1$\pm$0.8 (N=12) & strong (B) bars \\
  \noalign{\smallskip} 
length(bar)       & 4.1$\pm$0.4 kpc (N=62) & non-multibar galaxies \\
length(bar)       & 4.4$\pm$0.4 kpc (N=31) & multibar galaxies \\
length(nb)        & 0.7$\pm$0.3 kpc (N=43) & all nb  \\
length(nb)        & 0.6$\pm$0.2 kpc (N=8) & without main bar  \\
  \noalign{\smallskip} 
\hline
\end{tabular}
\end{center}
\end{table}

\clearpage
\newpage

\begin{figure}
\includegraphics[width=140mm]{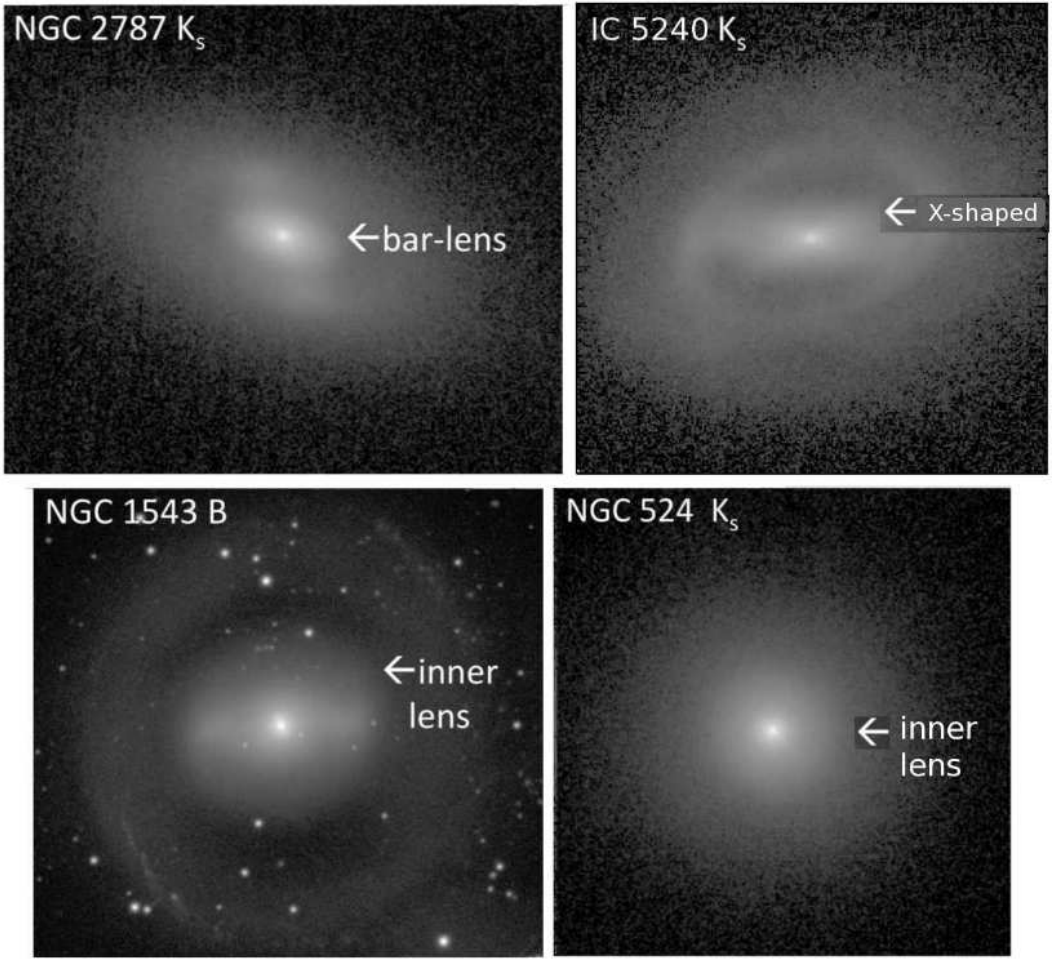}
\caption{Galaxies from the NIRS0S Atlas (Laurikainen et al. 2011)
demonstrating the different structure components. The images typically
reach a surface brightness of 23.5 mag arcsec$^{-2}$ in the $K_s$-band,
corresponding to a surface brightness of 27.5 mag
arcsec$^{-2}$ in the $B$-band. The galaxies are shown in a magnitude
scale. }
\label{figure-1}
\end{figure}
\clearpage
\newpage

\begin{figure}
\includegraphics[width=100mm]{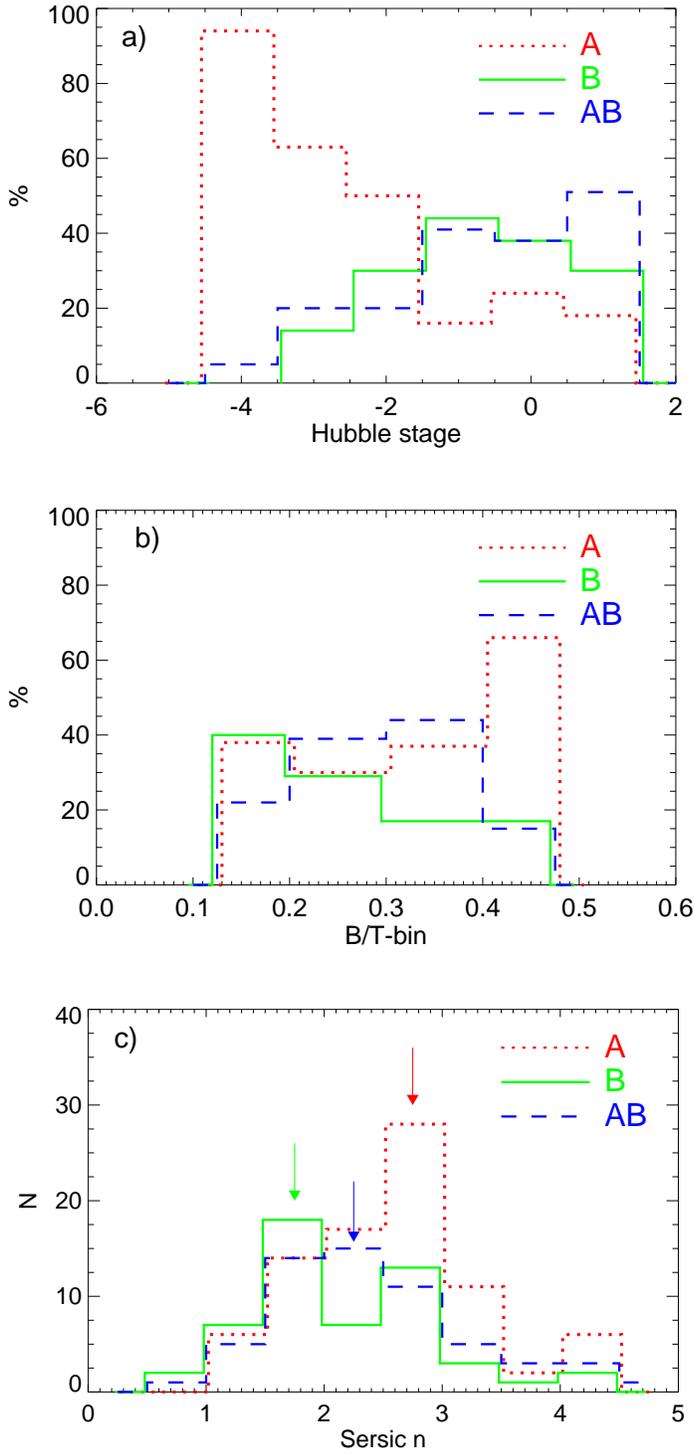}
\caption{ Galaxy fractions of different family classes (A,
AB, B) are shown as a function of Hubble type (a), bulge-to-total
($B/T$) flux-ratio (b), and the value of the S\'ersic index $n$ (c).
The statistics is based on the complete NIRS0S (include S0-Sa types).  The fractions given in
percentages are calculated with respect to the galaxy number in each
$B/T$ or Hubble type bin. The same is done in all following similar figures.
 }
\label{figure-2}
\end{figure}
\clearpage
\newpage

\begin{figure}
\includegraphics[width=100mm]{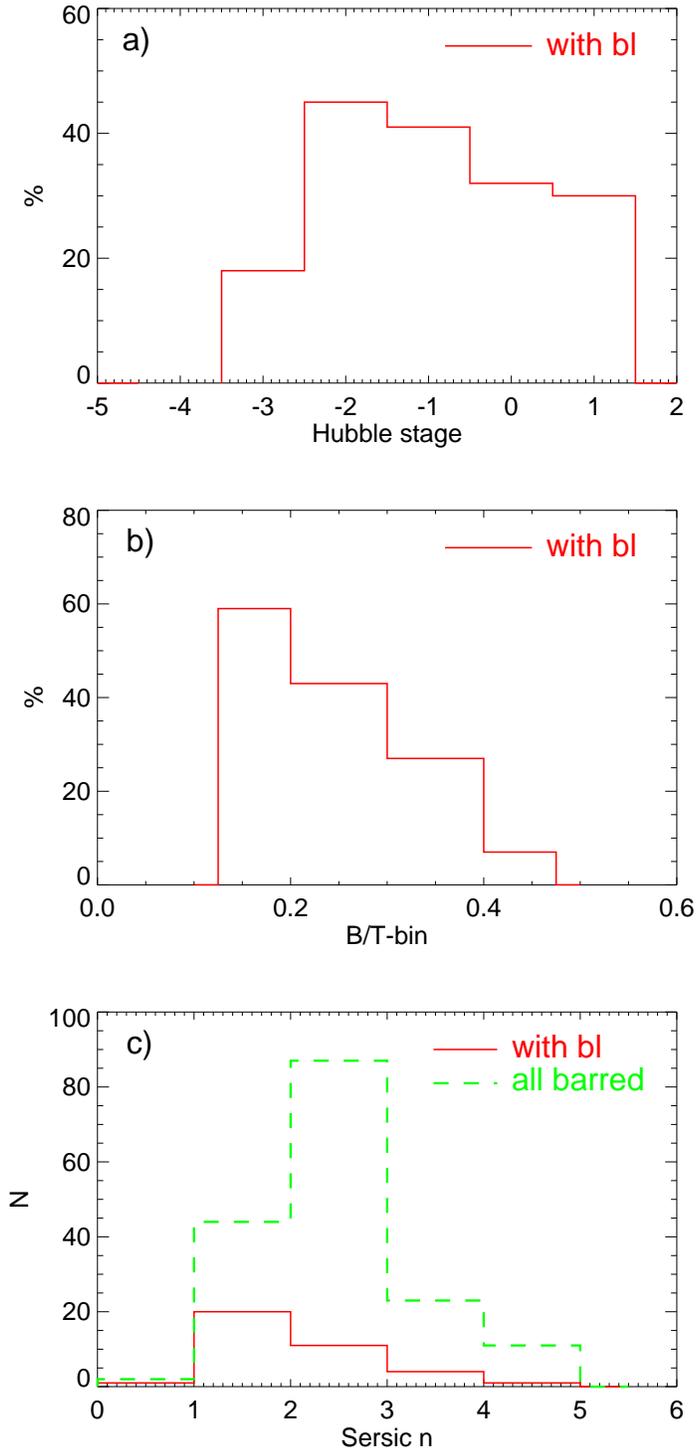}
\caption{Fractions of barred galaxies with barlenses are shown as a function
of Hubble type (a), and $B/T$ flux-ratio (b).  In (c)
the distribution of the S\'ersic index $n$ is shown for the galaxies
with barlenses, compared with that for all barred galaxies in
NIRS0S. }
\label{figure-3}
\end{figure}
\clearpage
\newpage

\begin{figure}
\includegraphics[width=160mm]{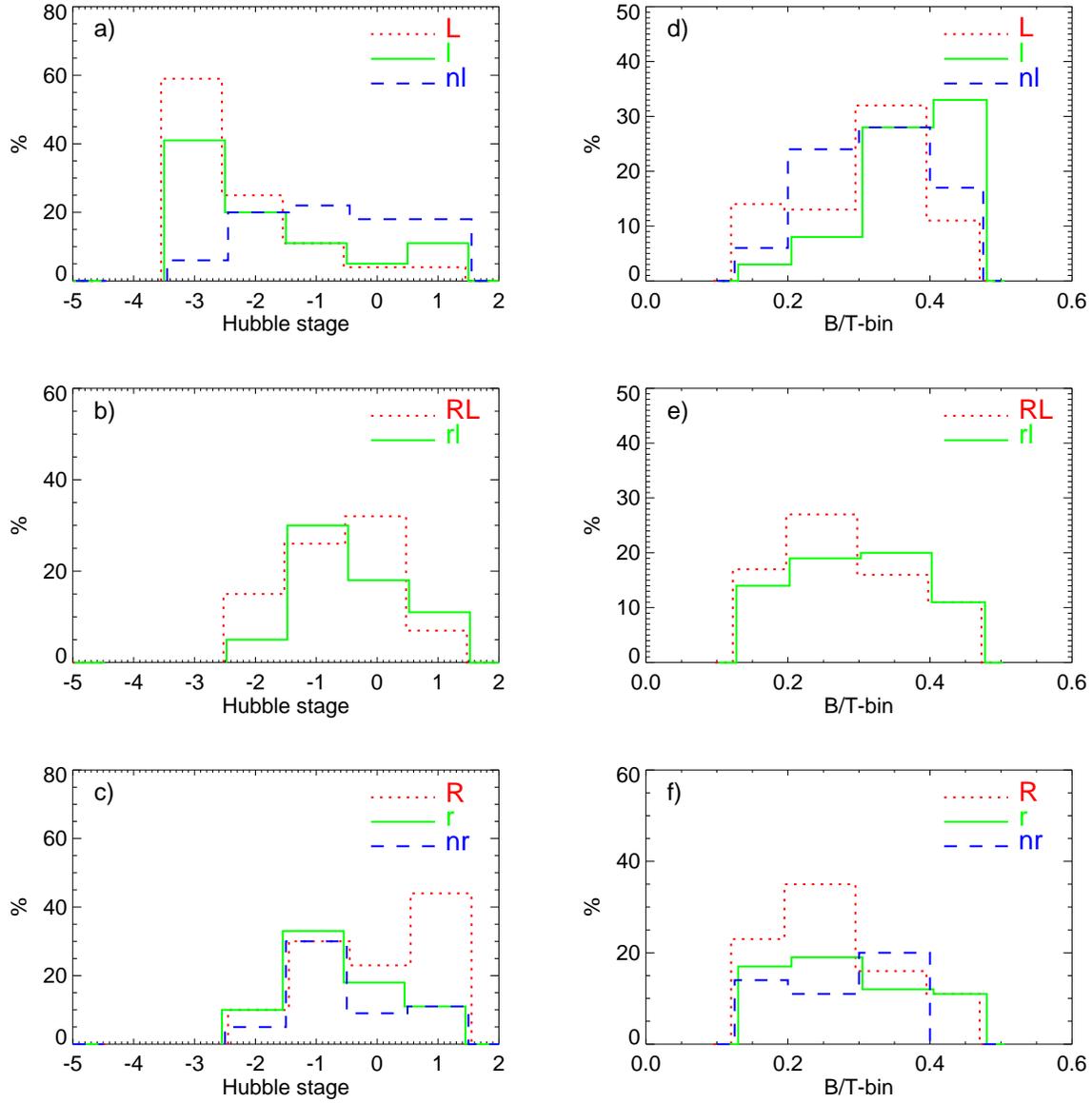}
\caption{ Fractions of the different structure components in barred
galaxies are shown, as a function of Hubble type (a)-(c), and as a
function of $B/T$ flux-ratio (d)-(f). The identifications of the
structures are from NIRS0S Atlas (the types rl and RL include also their
subtypes r'l and R'L). }
\label{figure-4}
\end{figure}
\clearpage
\newpage

\begin{figure}
\includegraphics[width=100mm]{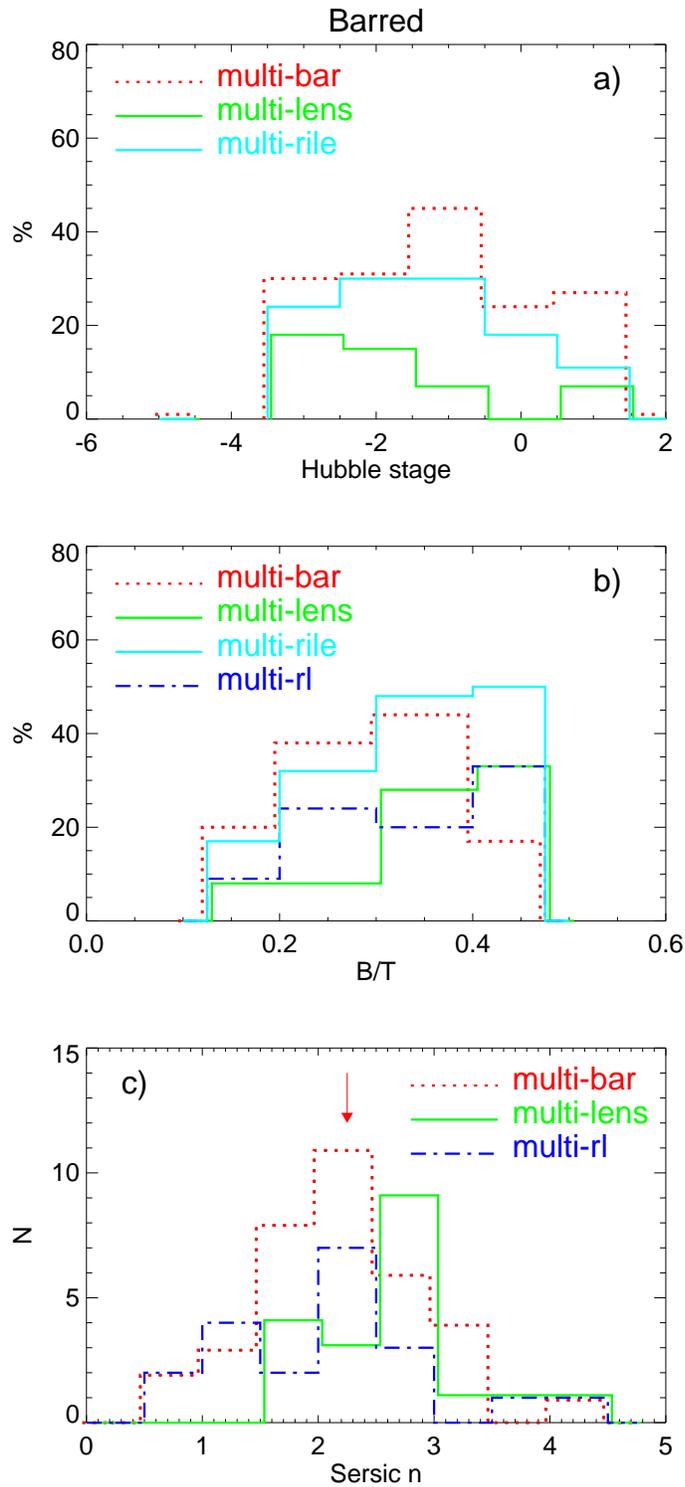}
\caption{Fractions of multiple components in barred galaxies:
(a) as a function of Hubble type, (b) $B/T$ flux-ratio, (c) and the
values of S\'ersic index. In the panels {\it multi-bar} = multiple bars,
{\it multi-lens} = multiple lenses in which both lenses are fully developed
(l,L), {\it multi-rl} = multiple ringlenses (rl,RL), {\it multi-rile} = multiple
lenses in which one of the lenses is a ringlens
(l,L,rl,RL). }
\label{figure-5}
\end{figure}
\clearpage
\newpage

\begin{figure}
\includegraphics[width=160mm]{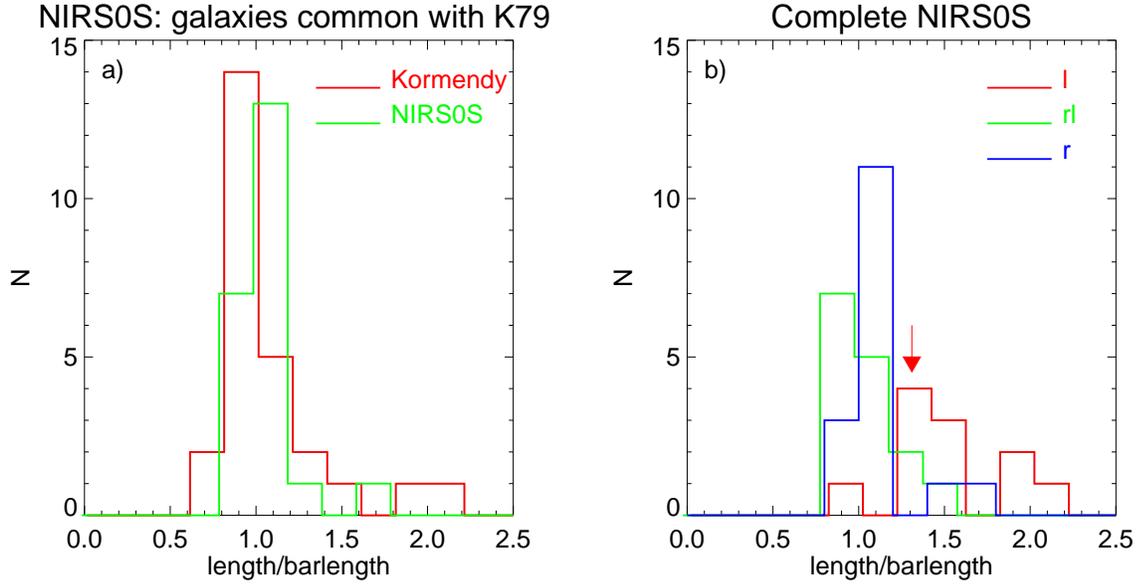}
\caption{{\it (a)} The lengths of the inner components (rings + ringlenses + lenses),
normalized to barlength: 41 galaxies common in NIRS0S and
in the sample by Kormendy (1979) are compared. 
{\it (b)} The complete NIRS0S is used, 
showing the lengths separately for the rings (r), ringlenses (rl) and lenses (l)
(see the details in Section 4.2). The arrow indicates the peak value for
the lens distribution. }
\label{figure-6}
\end{figure}

\begin{figure}
\includegraphics[width=160mm]{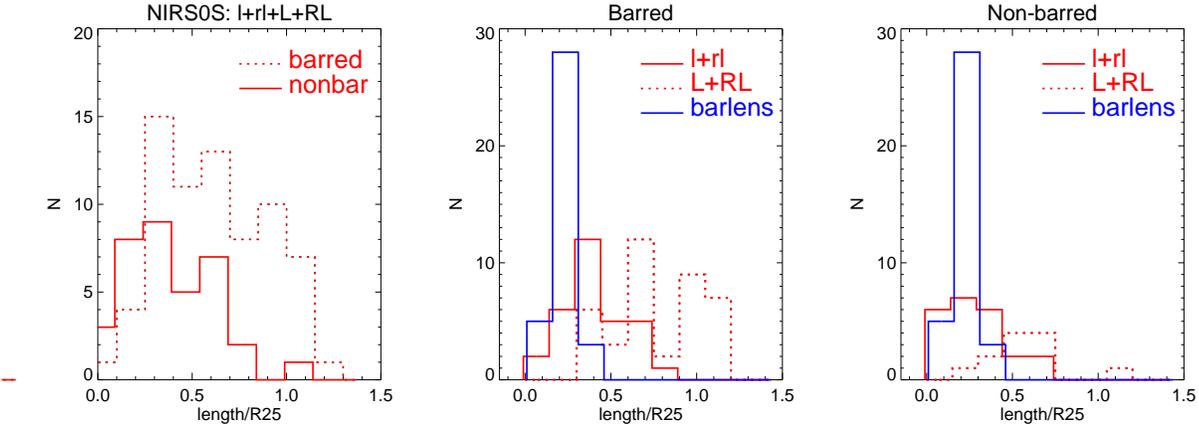}
\caption{The lengths of the lenses (including ringlenses) in barred
and non-barred galaxies in NIRS0S are compared. The sizes are normalized
to the galaxy size, R$_{25}$, given by the radius at the surface brightness of
25 magnitudes arcsec$^{-2}$ in the $B$-band, taken from RC3. {\it (a)}
The sizes of lenses in the barred and non-barred galaxies are shown, including both
the inner and outer lenses.  In the two other panels the inner 
lenses are shown separately for barred (b) and 
non-barred (c) galaxies. In these panels the length distribution 
for the barlenses is also shown: to facilitate comparison with inner lenses
in non-barred galaxies, it is shown also in c). }
\label{figure-7}
\end{figure}
\clearpage
\newpage

\begin{figure}
\includegraphics[width=140mm]{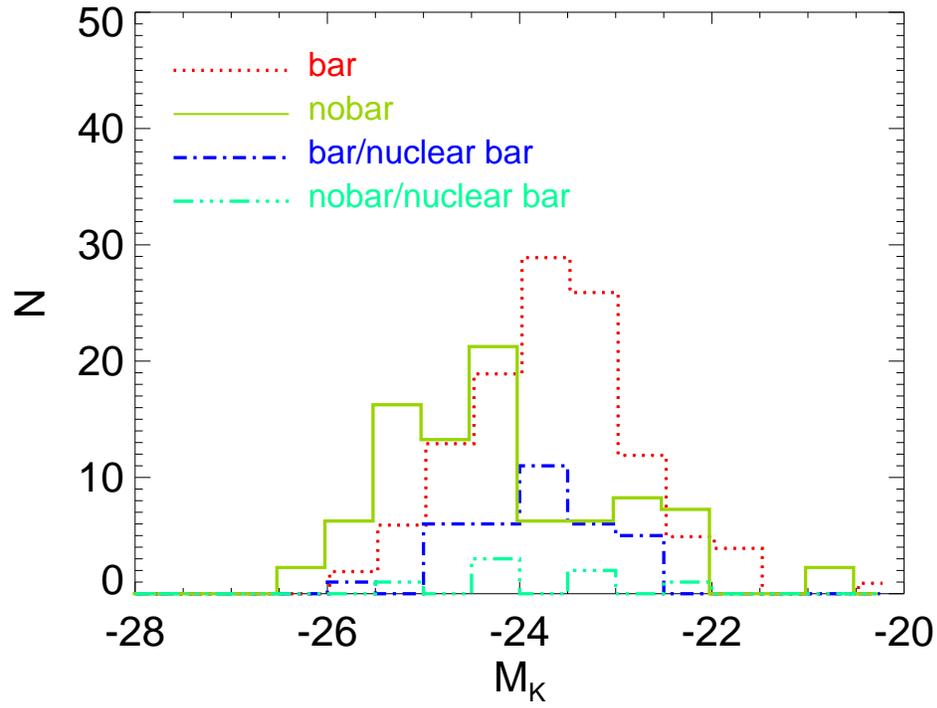}
\caption{The number histograms for the barred and the non-barred
galaxies, as well as for the nuclear bars, as a function of the absolute 
galaxy brightness in the $K$-band. As
explained in the text (Section 3) 2MASS extended apparent magnitudes
in the $K$-band were used, corrected for Galactic extinction. 
The distances are from the catalog by Tully (1988), using the Hubble
constant of H$_0$=75 km s$^{-1}$ Mpc$^{-1}$. }
\label{figure-8}
\end{figure}
\clearpage
\newpage

\begin{figure}
\includegraphics[width=140mm]{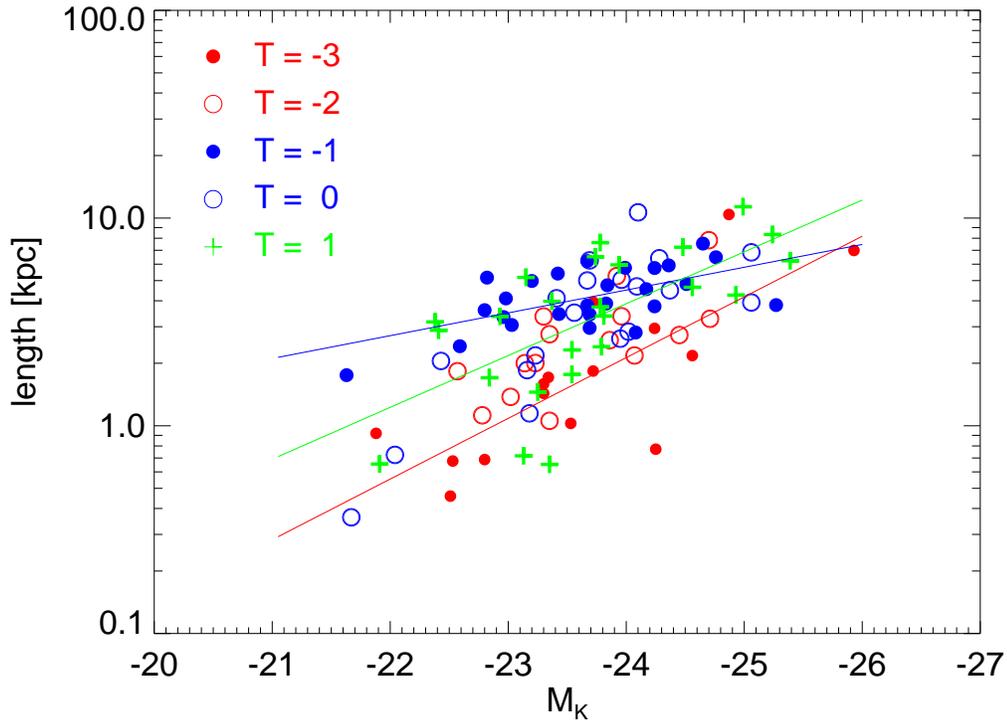}
\caption{The lengths of bars, inner rings, ringlenses, and
rings are shown as function of galaxy brightness.
The lengths are given in kiloparsecs. The lines are the
least square fits to the data points in the sub-categories T=$-$3 (red), T=$-$1 (blue), and T=1 (green).} 
\label{figure-9}
\end{figure}
\clearpage
\newpage

\begin{figure}
\includegraphics[width=140mm]{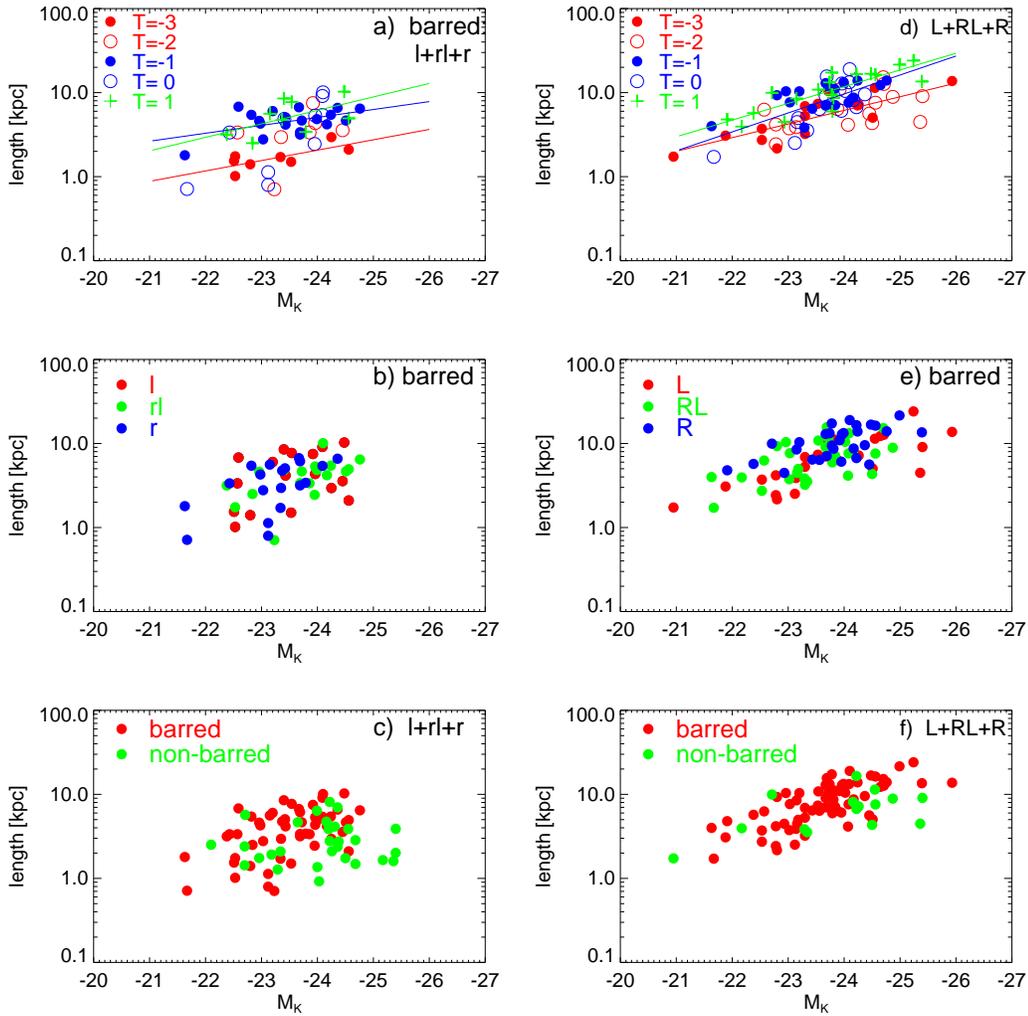}
\caption{The lengths of outer rings, ringlenses, and
rings are shown as function of galaxy brightness.
The lengths are given in kiloparsecs. }
\label{figure-10}
\end{figure}
\clearpage
\newpage

\begin{figure}
\includegraphics[width=160mm]{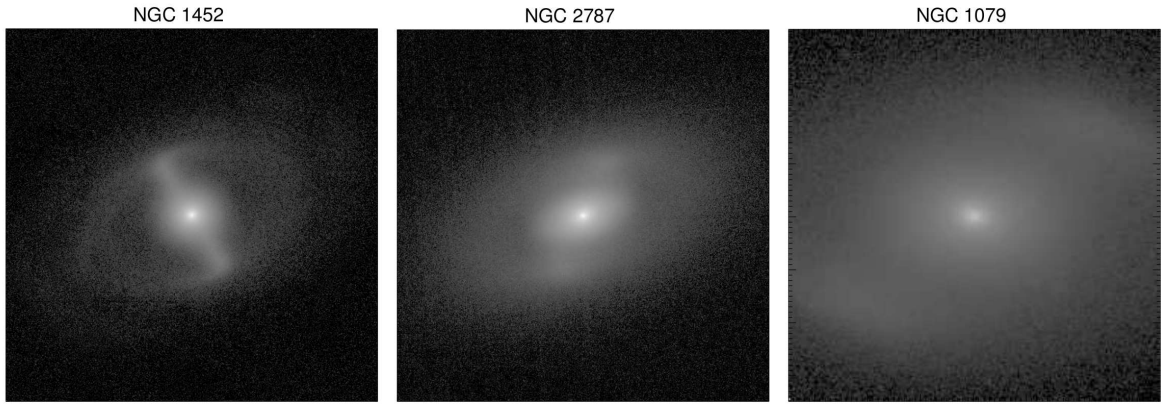}
\caption{Examples of barred galaxies: (a) NGC 1452 has as classical bar + ring,
(b) in NGC 2787 most of the light of the bar is concentrated to ansae,
and (c) in NGC 1709 the ansae of the bar look like starting to dissolve into the ring. }
\label{figure-11}
\end{figure}
\clearpage
\newpage

\begin{figure}
\includegraphics[width=100mm]{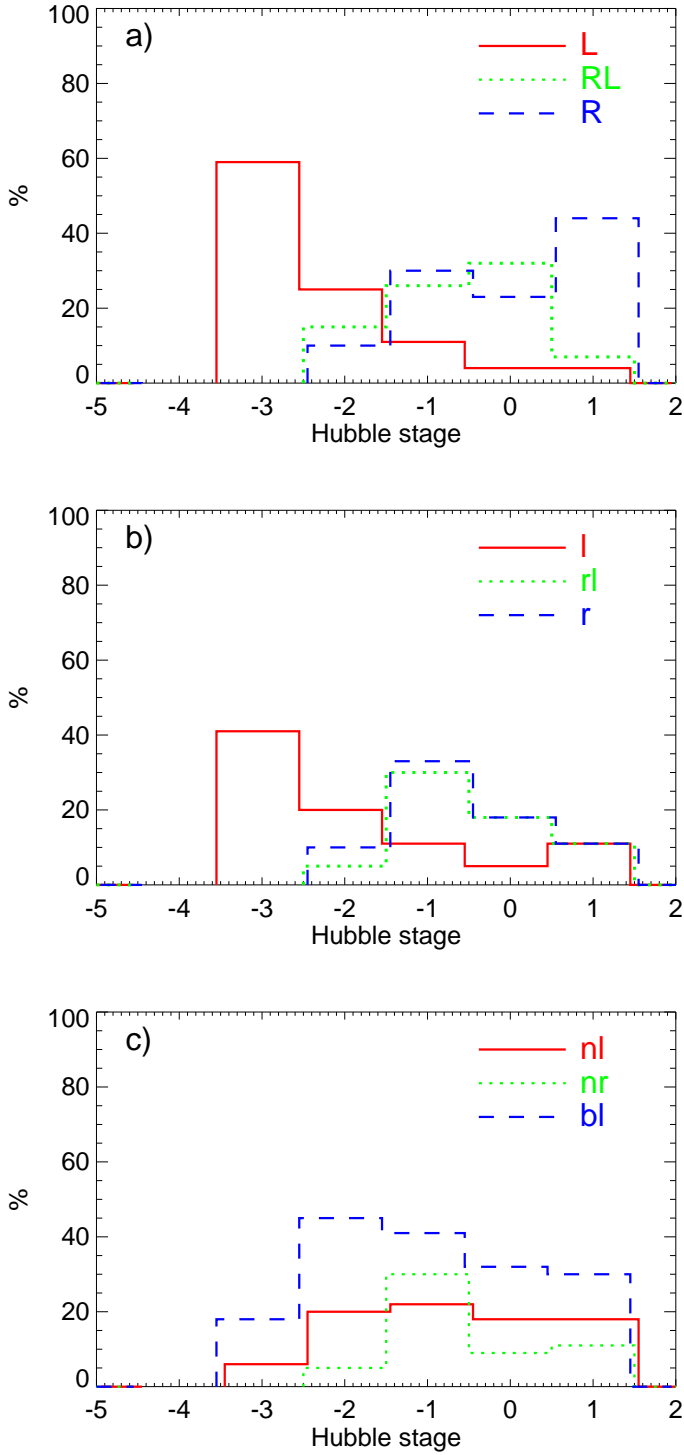}
\caption{Fractions of the outer, inner and nuclear features 
are shown as a function of Hubble type. The different
panels show the outer (a), inner (b) and nuclear (c) components. }
\label{figure-12}
\end{figure}
\clearpage
\newpage

\begin{figure}
\includegraphics[width=140mm]{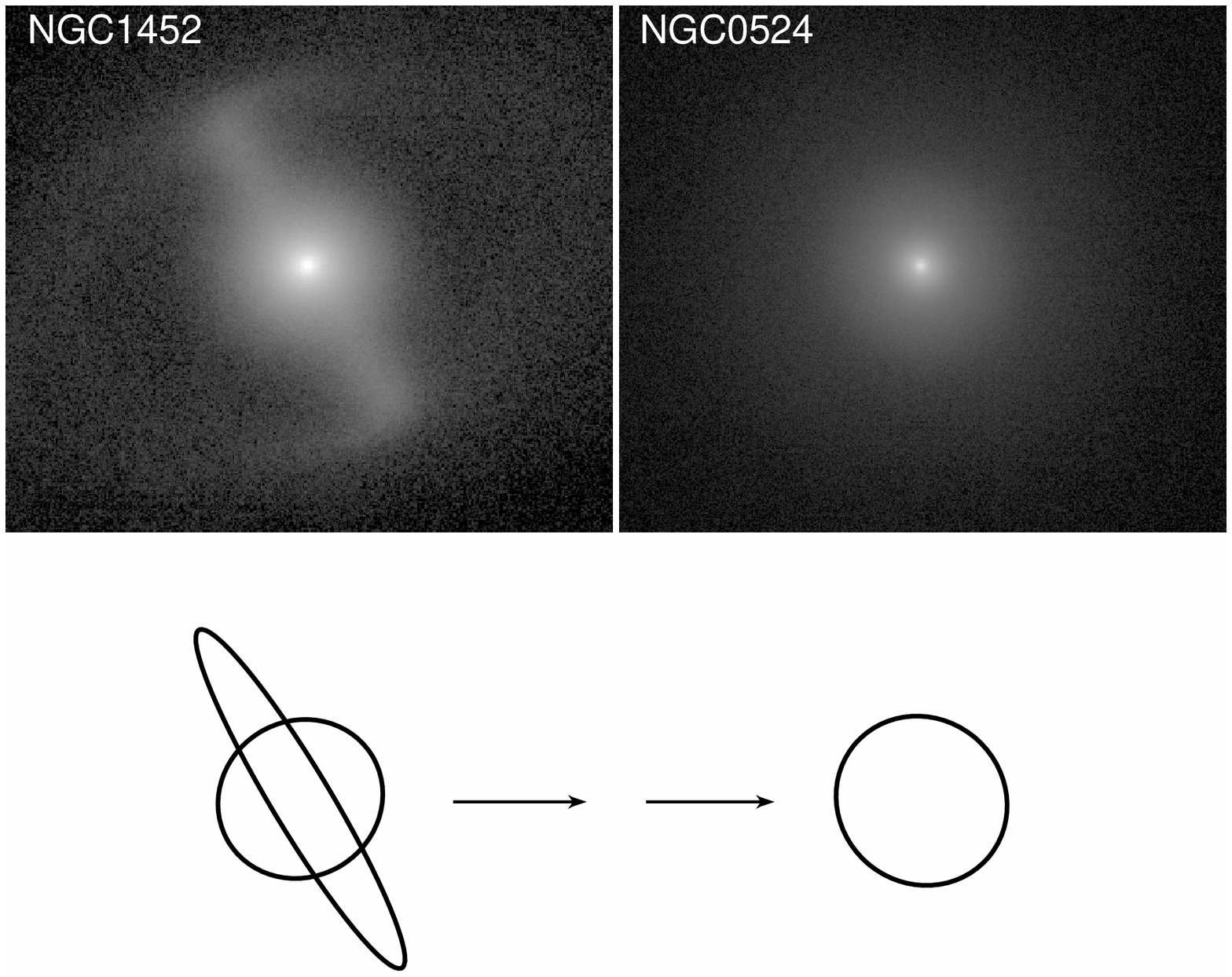}
\caption{A tentative example illustrating a possible bar dissolution process.
{\it Upper left panel:} NGC 1452 has a bar and a barlens
inside the bar. {\it Upper right panel:} NGC 524 has nuclear, inner and outer lenses,
but no bar.
{\it Lower panels:} a schematic illustration of the dissolution process. 
In the left panel the elongated component and the circle are shown in the
same scale as the bar and the barlens in NGC 1452. In the right panel
the circle corresponds to the inner lens of NGC 524, drawn in the right scale. }
\label{figure-13}
\end{figure}

\end{document}